\documentclass[aps,twocolumn,pra,showpacs,amsmath,amssymb,showkeys,10pt]{revtex4-1}

\usepackage{graphicx}
\usepackage{epstopdf}
\usepackage{braket}
\usepackage{hyperref}
\allowdisplaybreaks

\begin{document}

\title{Collective atomic correlations in absorptive optical bistability without adiabatic elimination: exemplifying nonclassicality from a linearized treatment of fluctuations}

\author{Th. K. Mavrogordatos}
\email[Email address: ]{themis.mavrogordatos@fysik.su.se}
\affiliation{Department of Physics, Stockholm University, SE-106 91, Stockholm, Sweden}

\date{\today}

\begin{abstract}
We determine the incoherent spectrum, squeezing properties and second-order correlation function of the collective atomic degrees of freedom in absorptive optical bistability. This is accomplished via a linearized Fokker-Planck equation in the positive $P$ representation, guided by the analysis of [H. J. Carmichael, Phys. Rev. A {\bf 33}, 3262 (1986)] which does not resort to adiabatic elimination. We focus on the regimes of weak and strong intracavity excitation, addressing the good-cavity and bad-cavity limits as well as the limit of collective strong coupling. Adiabatic elimination of the intracavity field sustained by an auxiliary resonator coupled to the ensemble is used to probe the atomic correlations via the formation of a collective emission channel. We compare to the corresponding expressions for the forwards-scattered light with reference to experimental results, discussing key differences between the lower and upper branch of the steady-state semiclassical bistability curve. Our analysis is carried out around the stable states situated far away from the turning points, where analytical expressions can be obtained self-consistently, demonstrating a clear departure from classical behavior. 
\end{abstract}


\maketitle

\section{Introduction}

Absorptive optical bistability was theoretically predicted in the late 1960s as a nonlinear phenomenon occurring in optical resonators containing saturable absorbers, while an experimental double-cavity arrangement was simultaneously proposed to measure the associated hysteresis \cite{Szoke1969}. Following a series of experimental and theoretical investigations (see \cite{LugiatoReview} for a comprehensive review on the topic), an exact analytical theory was reported about a decade later for the case of a unidirectional ring cavity where exact conditions for absorptive optical bistability were established \cite{Bonifacio1978Ring}. Since then, optical bistability has closely followed the development of quantum optics exemplifying several departures from classicality demonstrated by incoherent spectra, spectra of squeezing, and intensity correlation functions through extensive theoretical and experimental investigations. An important connection to superradiance was noticed in the 1970s; applying the spectral theorem for thermodynamic Green's functions in the low-temperature limit of the Dicke model produces the state equation of absorptive optical bistability \cite{Bowden1979}. The authors of this paper then proposed that ``the existence of the second-order superradiant phase transition in thermodynamic equilibrium and the existence of optical bistability as a first-order phase transition should stem from the same basic matter-light interaction''. 

The bad-cavity limit, in which the photon loss rate exceeds all other coupling rates, presents a connection between absorptive optical bistability and the quantum statistical treatment of superradiance (see e.g., \cite{Bonifacio1971I, Bonifacio1971II, Clemens2002}) and the spontaneous emission enhancement in the perturbative treatment of cavity QED (for the single-atom case see \cite{RiceCarmichaelIEEE}). In contrast, the correlation between unlike atoms vanishes in the good-cavity limit since the ``communication bandwidth'' of the mediating channel, being equal to the photon loss rate, tends to zero [see \cite{Carmichael1986} and Sec. 15.2.4 of \cite{QO2}]. In 1978, the authors of Ref. \cite{Bonifacio1978} employed a mean-field treatment to suggest a transition between a spectrum which is ``single-peaked when the system is in the cooperative stationary state'' and a ``dynamical Stark shift'' (or ``one-atom stationary state'') when the Rabi frequency of the incident field exceeds  the so-called {\it cooperative linewidth} of superfluorescence, set by a critical value of the atomic density, in the bad-cavity limit. The same authors in that year demonstrated a double-peaked Glauber-Sudarshan $P$ distribution in the steady state \cite{Bonifacio1978PhysRevLett}, and, alongside Agarwal and coworkers, reinforced the idea of a ``discontinuous formation of sidebands along the high-transmission branch'' of absorptive bistability. The latter employed a system-size expansion to take into account atomic correlation functions after having adiabatically eliminated the intracavity field \cite{Agarwal1978}; in doing so, they also obtained an asymptotic expression for the fluorescent spectrum in the ``limit of zero density''. An extensive summary and comparison of these early results relying on adiabatic elimination was given in \cite{Lugiato1979}. The distinction between like and unlike atom correlations, however, leads to continuous sideband formation in the fluorescent spectrum along the upper branch of bistability in the bad-cavity limit while there is no linewidth narrowing for the central peak \cite{Carmichael1981FS}. A number of the findings reported in the 1970s were corroborated and extended in 1983 by Carmichael and coworkers in \cite{Carmichael1983}, who derived expressions for the spectrum of fluctuations from the equations of motion of the covariance matrix in the positive $P$ representation within a linearized treatment; their analysis reaches out to dispersive bistability. In the dispersive limit, a connection between optical bistability and the nonlinear polarizability model had already been established in \cite{Drummond1980}. Three years lapsed to see the development of a ``linearized theory of fluctuations for absorptive bistability without adiabatic elimination of the atoms or the field'' in \cite{Carmichael1986}, invoking the notion of ``vacuum Rabi splitting'', as introduced in \cite{Sanchez1983, Agarwal1984}, to explain the oscillatory behavior of the second-order correlation function of the transmitted light. This feature was observed in the experiment of Raizen and coworkers \cite{Raizen1989} for a large cooperativity parameter and $\xi <1 $ such that a photon emitted by the atomic ensemble is likely to be captured and re-emitted multiple times before its escape from the cavity. In a similar approach revolving around linearization of the Fokker-Planck equation (FPE) for the Glauber-Sudarshan $P$ quasi-probability distribution, the spectrum of squeezing has been studied in \cite{Castelli1988}.

In the decades that followed, several experiments in quantum electrodynamics reached operating conditions of cavity QED, whose interest lies primarily with strong coupling for single atoms where the linearized treatment of fluctuations is no longer valid. Single-atom absorptive optical bistability was theoretically predicted in \cite{Savage1988} while spontaneous symmetry breaking and dressed-state polarization was associated with the so-called {\it zero-system size} of absorptive bistability in \cite{Alsing1991}, a critical effect experimentally demonstrated in \cite{Armen2009}. A few years later, photon antibunching was observed for a small number of two-level atoms strongly coupled to a high-finesse cavity \cite{Rempe1991}, while a violation of Schwarz inequality due to nonclassical correlations in the bunched light emitted from a collection of atoms strongly coupled to a cavity mode was reported in \cite{Mielke1998}. Experimental data at weak excitation were also presented in \cite{Foster2000}. Closer to our days, atom-light field correlations for a small number of atoms strongly coupled to the intracavity field in optical bistability were discussed in \cite{Dombi2013}, where it was reported that the size of these correlations increases with the number of atoms for strong driving, along the upper branch of the bistability curve. The authors bring up the similarity of the pronounced correlations with the ``cavity-QED limit'' of a thresholdless device and the associated gradual evolution of a quantum nonlinear process, as opposed to the conventional ``thermodynamic limit'' where one sends the saturation photon number to infinity \cite{Rice1994}. Alternatively, the Bogolyubov-Born-Green-Kirkwood-Yvon hierarchy had been used in \cite{Gladush2011} to study the scattering of resonant radiation in a dense two-level medium, where dispersive bistability is regarded as a switching mechanism between different spectral patterns.

After the achievement of strong coupling of a Bose Einstein condensate (BEC) to an ultra-high finesse resonator in \cite{Brennecke2007} and at a single-atom level \cite{Colombe2007} \textemdash{defining} a conceptually new regime of cavity QED \textemdash{and} the recent theoretical proposals to control optical bistability for a configuration comprising a BEC in a cavity \cite{Yang2011, Dalafi2013, Rosanov2006} akin to the function of an optical switch, the quantum correlations of the field scattered by the intracavity atoms are not only within experimental reach but also in position to signal criticality arising from underlying nonlinear interactions. A hybrid optomechanical setup to tune bistability has been used in \cite{Yasir2015} where a BEC is trapped in a Fabry-P\'{e}rot resonator with a movable mirror. At the same time, active environments have been very recently shown to enhance the system nonlinearity in the strong-coupling regime of cavity QED \cite{Hagenmuller2020, Schutz2020}. Our interest in this work is with the linearized treatment of atomic polarization fluctuations without resorting to adiabatic elimination. Light-matter correlation functions in the weak-excitation limit of absorptive bistability have been recently presented in \cite{Mavrogordatos2020} within the same framework. The procedure requires a large number of atoms and calls for an experimental investigation for the applicability of the derived formulas when this number ceases to be large in light of strong-coupling conditions for single atoms; some restrictions are immediately obvious from general requirements such as the non-negativity of the intensity correlation function or arise from the comparison with alternative theoretical approaches, such as the pure-state factorization for weak excitation \cite{Carmichael1991OC}. We remain within purely absorptive optical bistability and apply a linearization procedure about the steady states lying along the very beginning of the cooperative branch and at the high-excitation part of the input-output curve; fluctuations are always assumed to follow a Gaussian distribution. After delineating the transition from the master equation (ME) to the linearized FPE for the fluctuations in the phase space in Sec. \ref{sec:MEtoFPE}, we proceed to the study of atomic fluctuations about the steady state in Sec. \ref{sec:corrincohspectr}. We derive an approximate formula for the correlation spectrum in the weak-excitation limit in Subsec. \ref{subsec:spewel}, placing particular emphasis on the bad and good cavity limits as well as on the limit of collective strong coupling between the atomic ensemble and the intracavity field. We then move on to consider the high-excitation part of the upper branch in Subsec. \ref{subsec:SEUB} giving approximate expressions for the incoherent spectrum of the forward and side-scattered fields. A proposal for accessing the collective atomic emission is formulated in Sec. \ref{sec:accessatem}, taking into account that side scattering does not occur via a single collective mode. After a short discussion on squeezing in Sec. \ref{sec:sqafl}, we calculate the second-order correlation function of the collective atomic polarization as a system degree of freedom in both weak and strong-excitation limits of bistability in Sec. \ref{sec:2ndorderC}, before summarizing our findings and bringing up the relevance to quantum optical experiments in Sec. \ref{sec:conclusions}.  

\section{Constructing the Fokker-Planck equation for fluctuations in absorptive optical bistability}
\label{sec:MEtoFPE}

We begin with the {\it master equation (ME) of optical bistability} for a collection of $N$ homogeneously broadened two-level atoms on resonance with a single cavity mode coherently driven with amplitude $\bar{\mathcal{E}}_0$, and subject to radiative damping \cite{CarmichaelSatchell1986},
\begin{align}\label{eq:MEbist}
\frac{d\rho}{dt}&=-i\frac{1}{2}\omega_0 [J_z, \rho]-i\omega_0[a^{\dagger}a, \rho] \notag \\
&+g[a^{\dagger}J_{-}-aJ_{+},\rho]-i[\bar{\mathcal{E}}_0 e^{-i\omega_0 t}a^{\dagger}+\bar{\mathcal{E}}_0^{*} e^{i\omega_0 t}a, \rho] \notag \\
&+\frac{\gamma}{2}\left(\sum_{j=1}^{N}2\sigma_{j-}\rho \sigma_{j+}-\frac{1}{2}J_{z}\rho - \frac{1}{2}\rho J_{z} -N\rho\right)\notag \\
&+\kappa(2a\rho a^{\dagger}-a^{\dagger}a\rho-\rho a^{\dagger}a),
\end{align}
where $\rho$ is the system density operator, $a^{\dagger}$ and $a$ are the creation and annihilation operators, respectively, for the cavity photons, and $J_{-}$, $J_{+}$ and $J_{z}$ are collective atomic operators defined via the pseudospin operators $\sigma_{j \pm},\sigma_{j z}$, $j=1,\ldots, N$, for each of the $N$ individual atoms comprising the collection,
\begin{equation}\label{eq:atomopdef}
J_{\pm} \equiv \sum_{j=1}^{N}\sigma_{j \pm}, \quad J_z \equiv \sum_{j=1}^{N}\sigma_{jz},
\end{equation}
satisfying the familiar commutation relations of angular momentum operators,
\begin{equation}\label{eq:commatomop}
[J_{+}, J_{-}]=J_{z}, \quad [J_{\pm}, J_{z}]= \mp 2 J_{\pm}.
\end{equation}
As far as the energy scale of the dynamics is concerned, $\omega_0$ is the resonant frequency coinciding with the frequency of the drive, $\gamma$ is the atomic decay rate, $\kappa$ is the decay rate for the cavity field, and $g$ is the atom-field dipole coupling strength, which is assumed to have the same value for every atom, $g=[\omega_0 d^2/(2 \hbar \epsilon_0 V_Q)]^{1/2}$, where $d$ is the atomic dipole moment, $\epsilon_0$ is the vacuum permittivity and $V_Q$ is the mode volume. We also take $\overline{n}=0$ for the complete reservoir coupled to the bistable absorber, which is appropriate at optical frequencies. In ME \eqref{eq:MEbist}, we have also neglected the inhomogeneous broadening caused by the random movement of atoms comprising the ensemble (Doppler broadening), which is appreciable at room temperature. 

Our next step is the construction of a phase-space representation which is equivalent to the ME \eqref{eq:MEbist} within a linearized treatment of quantum fluctuations. In the positive $P$ representation, with its initial formulation extending Haken's theory for the laser \cite{HakenBook}, we employ a ten-dimensional space with five independent phase-space variables $(\alpha, \alpha_{*}, v, v_{*},m)$  featuring in the FPE constructed from the ME \eqref{eq:MEbist}. One first defines a suitable normally-ordered characteristic function 
\begin{equation}\label{eq:chfun}
\chi_N(\zeta,\zeta_{*},\eta, \eta_{*},\theta) \equiv {\rm tr}(\rho e^{i\zeta_{*}a^{\dagger}} e^{i\zeta a} e^{i\eta_{*}J_{+}}e^{i\theta J_z} e^{i\eta J_{-}}),
\end{equation}
generating the hierarchy of normal-ordered operator averages, and then the quasi-distribution function \cite{CarmichaelSatchell1986, Carmichael1983}
\begin{equation}\label{eq:P}
\begin{aligned}
&\chi_N(\zeta,\zeta_{*},\eta,\eta_{*},\theta) \\
& \equiv \int d^2 \alpha \int d^2\alpha_{*} \int d^2v \int d^2v_{*} \int d^2 m \\
& P(\alpha, \alpha_{*},v,v_{*},m)e^{-i\zeta_{*}\alpha_{*}}e^{-i\zeta\alpha} e^{-i\eta_{*}v_{*}} e^{-i\eta v} e^{-i\theta m},
\end{aligned}
\end{equation}
where each integration extends over the entire complex plane; it is important to note that the characteristic function $\chi_N(\zeta,\zeta_{*},\eta, \eta_{*},\theta)$ and the quasi-distribution function $P(\alpha, \alpha_{*},v,v_{*},m)$ do {\it not} form a Fourier-transform pair unlike what happens in the Glauber-Sudarshan $P$ representation \cite{Drummond1980PositiveP, QO2}. Physical realizations have ensemble averages satisfying $(\overline{\alpha_{*}})_{P}=(\overline{\alpha^{*}})_{P}$, $(\overline{v_{*}})_{P}=(\overline{v^{*}})_{P}$ and $(\overline{m})_{P}=(\overline{m^{*}})_{P}$ (in this instance exclusively the long bar signifies a statistical average in contrast to the short bar on top of a variable or operator reserved for scaling further on). For the phase-space variables we adopt the scaling [see Sec. 15.2.1 of \cite{QO2}]
\begin{subequations}\label{eq:scalingFP}
\begin{align}
i e^{-i\phi_0}\alpha&=n_{\rm sc}^{1/2}\bar{\alpha}, \label{eq:scalingFP_a}\\
-i e^{i\phi_0}\alpha_{*}&=n_{\rm sc}^{1/2}\bar{\alpha}_{*}, \label{eq:scalingFP_b}\\
i \sqrt{2}\, e^{-i\phi_0}v&=N \bar{v}, \label{eq:scalingFP_c}\\
-i \sqrt{2}\, e^{i\phi_0}v_{*}&=N \bar{v}_{*}, \label{eq:scalingFP_d}\\
m&=N\bar{m}, \label{eq:scalingFP_e}
\end{align}
\end{subequations}   
in parallel with the operator relations
\begin{subequations}\label{eq:ScalingOp}
\begin{align}
i e^{-i \phi_0}a &=n_{\rm sc}^{1/2}\overline{a}, \label{eq:ScalingOp_a}\\
-i e^{i \phi_0} a^{\dagger} &=n_{\rm sc}^{1/2}\overline{a}^{\dagger}, \label{eq:ScalingOp_b}\\
i \sqrt{2}\, e^{-i \phi_0}J_{-} &=N \overline{J}_{-}, \label{eq:ScalingOp_c}\\
-i \sqrt{2}\, e^{i\phi_0}J_{+}&=N \overline{J}_{+}, \label{eq:ScalingOp_d}\\
J_{z}&=N\overline{J}_{z}. \label{eq:ScalingOp_e}
\end{align}
\end{subequations}   
where $n_{\rm sc}\equiv \gamma^2/(8g^2)$ is the saturation photon number associated with the weak-coupling limit of absorptive optical bistability, and the phase $\phi_0$ is introduced to ensure that the driving-field amplitude appears hereinafter as a real parameter. It is also convenient to transform to a frame rotating with the drive frequency $\omega_0$,
\begin{subequations}\label{eq:rotatingframe}
\begin{align}
&\bar{\alpha}=e^{-i\omega_0 t} \tilde{\bar{\alpha}}, \quad \bar{\alpha}^{*}=e^{i\omega_0 t} \tilde{\bar{\alpha}}^{*}, \label{eq:rotatingframe_a}\\
&\bar{v}=e^{-i\omega_0 t} \tilde{\bar{v}}, \quad \bar{v}^{*}=e^{i\omega_0 t} \tilde{\bar{v}}^{*}. \label{eq:rotatingframe_b}
\end{align}
\end{subequations}
Hereinafter, the tilde on top of the operators and phase-space variables signifies a transformation to a frame rotating by the resonant frequency $\omega_0$. At this point, we introduce fluctuations scaling with respect to the number of atoms such that the small-noise limit admitting the linearized treatment is attained for $N \gg 1$. Based on \cite{Carmichael1986} and Sec. 15.2.1 of \cite{QO2} we write
\begin{subequations}\label{eq:scaleN}
\begin{align}
\bar{\alpha}&=\braket{\bar{a}(t)} + N^{-1/2}z, \label{eq:scaleN_a}\\
\bar{\alpha}_{*}&=\braket{\bar{a}^{\dagger}(t)} + N^{-1/2}z_{*}, \label{eq:scaleN_b}\\
\bar{v}&=\braket{\bar{J}_{-}(t)} + N^{-1/2} \nu, \label{eq:scaleN_c}\\
\bar{v}_{*}&=\braket{\bar{J}_{+}(t)} + N^{-1/2} \nu_{*}, \label{eq:scaleN_d}\\
\bar{m}&=\braket{\bar{J}_{z}(t)} + N^{-1/2}\mu, \label{eq:scaleN_e}
\end{align}
\end{subequations}
and demanding that terms of order $N^{1/2}$ in the FPE vanish, yields the Maxwell-Bloch equations (macroscopic law)
\begin{subequations}\label{eq:MB}
\begin{align}
\kappa^{-1} \frac{d\braket{\tilde{\bar{a}}}}{dt}&=-\braket{\tilde{\bar{a}}} + 2C \braket{\tilde{\bar{J}}_{-}} + Y, \label{eq:MBa}\\
\kappa^{-1} \frac{d\braket{\tilde{\bar{a}}^{\dagger}}}{dt}&=-\braket{\tilde{\bar{a}}^{\dagger}} + 2C \braket{\tilde{\bar{J}}_{+}} + Y, \label{eq:MBb}\\
\left(\frac{\gamma}{2}\right)^{-1}\frac{d\braket{\tilde{\bar{J}}_{-}}}{dt}&=-\braket{\tilde{\bar{J}}_{-}} + \braket{\bar{J}_{z}} \braket{\tilde{\bar{a}}}, \label{eq:MBc}\\
\left(\frac{\gamma}{2}\right)^{-1}\frac{d\braket{\tilde{\bar{J}}_{+}}}{dt}&=-\braket{\tilde{\bar{J}}_{+}} + \braket{\bar{J}_{z}} \braket{\tilde{\bar{a}}^{\dagger}}, \label{eq:MBd}\\
\gamma^{-1}\frac{d\braket{\bar{J}_{z}}}{dt}&=-(\braket{\bar{J}_z}+1)-\frac{1}{2}\left(\braket{\tilde{\bar{J}}_{+}}\braket{\tilde{\bar{a}}}+ \braket{\tilde{\bar{J}}_{-}}\braket{\tilde{\bar{a}}^{\dagger}}\right), \label{eq:MBe}
\end{align}
\end{subequations}
in which $Y \equiv n_{\rm sc}^{-1/2}(|\bar{\mathcal{E}_0}|/\kappa)$ is the scaled input. Their steady-state solution reads
\begin{subequations}\label{eq:MBss}
\begin{align}
&\braket{\tilde{\bar{a}}}_{\rm ss}=\braket{\tilde{\bar{a}}^{\dagger}}_{\rm ss} \equiv X, \label{eq:MBss_a}\\
&\braket{\tilde{\bar{J}}_{-}}_{\rm ss}=\braket{\tilde{\bar{J}}_{+}}_{\rm ss}=-\frac{X}{1 + X^2}, \label{eq:MBss_b}\\
&\braket{\bar{J}_{z}}_{\rm ss}=-\frac{1}{1+X^2}, \label{eq:MBss_c}
\end{align}
\end{subequations}
where the intracavity amplitude $X$ is determined by the state equation of absorptive optical bistability
\begin{equation}\label{eq:BistSE}
Y=X\left(1 + 2C \frac{1}{1+X^2}\right),
\end{equation}
where $2C=2Ng^2/(\kappa\gamma)$ is the {\it cooperativity parameter}. The derivative $dY/dX$ vanishes for $X_{\pm}^2=C-1\pm \sqrt{C(C-4)}$, and bistability occurs for $C>4$. In this work, we will focus on the linearization about the two stable steady states of the bistability curve, one along the {\it lower branch}, defined by the inequality $X<X_{-}$ and the other along the {\it upper branch}, defined by the inequality $X>X_{+}$.

Within the positive $P$ representation and a frame rotating with the frequency of the drive, we denote the vector of fluctuations by $\tilde{\boldsymbol{Z}} \equiv (\tilde{z}, \tilde{z}_{*}, \tilde{\nu}, \tilde{\nu}_{*}, \mu)$, where $(\tilde{z}, \tilde{z}_{*})$ and $(\tilde{\nu}, \tilde{\nu}_{*})$ are pairs of independent complex variables. Derivatives in the extended configuration space of ten variables are suitably defined so as to produce a positive semi-definite diffusion. The covariance matrix corresponding to the vector $\tilde{\boldsymbol{Z}}(t)$ is defined with respect to the scaled quasi-distribution function for the fluctuations
\begin{equation}
\begin{aligned}
&\tilde{\bar{P}}(\tilde{z}, \tilde{z}_{*}, \tilde{\nu}, \tilde{\nu}_{*},\mu)\equiv \frac{1}{2}n_{\rm sc}N^{1/2}\\
& \times P(\tilde{\bar{\alpha}}(\tilde{z},t), \tilde{\bar{\alpha}}_{*}(\tilde{z}_{*},t), \tilde{\bar{v}}(\tilde{\nu},t), \tilde{\bar{v}}_{*}(\tilde{\nu}_{*},t), m(\mu,t)).
\end{aligned}
\end{equation}
The function $\tilde{\bar{P}}(\tilde{z}, \tilde{z}_{*}, \tilde{\nu}, \tilde{\nu}_{*},\mu)$, in the positive $P$ representation, solves the linearized FPE [see Eq. (15.75) of \cite{QO2}]
\begin{equation}\label{eq:FPE}
\frac{\partial \tilde{\bar{P}}}{\partial t}=\left(-\tilde{\boldsymbol{Z}}^{\prime \top} \bar{\boldsymbol{J}}_{\rm ss} \tilde{\boldsymbol{Z}} + \frac{1}{2}\tilde{\boldsymbol{Z}}^{\prime \top} \bar{\boldsymbol{D}}_{\rm ss} \tilde{\boldsymbol{Z}}^{\prime} \right) \tilde{\bar{P}},
\end{equation}
with
\begin{equation}
\tilde{\boldsymbol{Z}} \equiv \begin{pmatrix}
\tilde{z} \\ \tilde{z}_{*} \\ \tilde{\nu} \\ \tilde{\nu}_{*} \\ \tilde{\mu}
\end{pmatrix}, \quad \quad \tilde{\boldsymbol{Z}}^{\prime} \equiv \begin{pmatrix}
\partial / \partial\tilde{z} \\ \partial / \partial\tilde{z}_{*} \\ \partial / \partial\tilde{\nu} \\ \partial / \partial\tilde{\nu}_{*} \\ \partial / \partial\tilde{\mu}
\end{pmatrix}.
\end{equation}
The covariance matrix is then defined as
\begin{equation}\label{eq:defCovMatrix}
\boldsymbol{C}_{\rm ss}(\tau) \equiv \lim_{t \to \infty}\left(\overline{\tilde{\boldsymbol{Z}}(t)\tilde{\boldsymbol{Z}}^{\top}(t+\tau)}\right)_{\tilde{\bar{P}}},
\end{equation}
and obeys the equation of motion
\begin{equation}\label{eq:eomC}
\frac{d\boldsymbol{C}_{\rm ss}}{d\tau}=\begin{cases}\boldsymbol{C}_{\rm ss} \bar{\boldsymbol{J}}_{\rm ss}^{\top} \quad \quad \tau > 0 \\
\bar{\boldsymbol{J}}_{\rm ss}\boldsymbol{C}_{\rm ss} \quad  \quad \tau < 0 \end{cases},
\end{equation}
while at $\tau=0$,
\begin{equation}\label{eq:initialcondC}
\bar{\boldsymbol{J}}_{\rm ss} \boldsymbol{C}_{\infty} + \boldsymbol{C}_{\infty} \bar{\boldsymbol{J}}_{\rm ss} = -\bar{\boldsymbol{D}}_{\rm ss}.
\end{equation} 
In the linearized FPE for the fluctuations \eqref{eq:FPE} as well as in Eqs. \eqref{eq:eomC} and \eqref{eq:initialcondC}, $\bar{\boldsymbol{J}}_{\rm ss}$ is the Jacobian matrix and $\bar{\boldsymbol{D}}_{\rm ss}$ is the diffusion matrix. These two matrices assume the form
\begin{equation}\label{eq:JacobFull}
\bar{\boldsymbol{J}}_{\rm ss}=\frac{\gamma}{2}\begin{pmatrix}
-\xi & 0 & \xi 2C & 0 & 0 \\
0 & -\xi & 0 & \xi 2C & 0 \\
-1/(1+X^2) & 0 & -1 & 0 & X \\
0 & -1/(1+X^2) & 0 & -1 & X \\
X/(1+X^2) & X/(1+X^2) & -X & -X & -2
\end{pmatrix}
\end{equation}
and
\begin{equation}\label{eq:DiffM}
\bar{\boldsymbol{D}}_{\rm ss}=\gamma\frac{X^2}{1+X^2} \begin{pmatrix}
0 & 0 & 0 & 0 & 0\\
0 & 0 & 0 & 0 & 0\\
0 & 0 & -1 & 0 & 0\\
0 & 0 & 0 & 1 & 0\\
0 & 0 & 0 & 0 & 4
\end{pmatrix},
\end{equation}
respectively, with $\xi \equiv 2\kappa/\gamma$ the ratio of the two decay rates whose relation to $C$ defines distinct regions of operation for the bistable absorber. The diffusion matrix of Eq. \eqref{eq:DiffM} is manifestly nonpositive semidefinite. In the linearized theory of fluctuations under current consideration, however, steady-state moments and the spectrum of fluctuations can be calculated in the original space by a naive application of familiar formal expressions to a FPE in the Glauber-Sudarshan $P$-representation with a nonpositive-definite diffusion. In other words, we carry on with the calculations exactly as we would if the diffusion matrix were positive semidefinite. Of course, this is no longer possible when nonlinearity enters into play and divergent trajectories appear in the stochastic simulations of absorptive optical bistability within the positive $P$ representation \cite{CarmichaelSatchell1986}.

We can now introduce the set of system operators $(\bar{a}, \bar{a}^{\dagger}, \bar{J}_{-}, \bar{J}_{+}, \bar{J}_z)$, defined in correspondence with the scaling relations \eqref{eq:scalingFP}, and the fluctuation operators given by
\begin{subequations}\label{eq:floper}
\begin{align}
\Delta\overline{a} &\equiv \overline{a} - \braket{\overline{a}}_{\rm ss}, \label{eq:floper_a}\\
\Delta\overline{a}^{\dagger} &\equiv \overline{a}^{\dagger} - \braket{\overline{a}^{\dagger}}_{\rm ss}, \label{eq:floper_b}\\
\Delta \overline{J}_{-} &\equiv \overline{J}_{-} - \braket{\overline{J}_{-}}_{\rm ss} \label{eq:floper_c}\\
\Delta \overline{J}_{+} &\equiv \overline{J}_{+} - \braket{\overline{J}_{+}}_{\rm ss} \label{eq:floper_d}\\
\Delta \overline{J}_{z} &\equiv \overline{J}_z - \braket{\overline{J}_{z}}_{\rm ss}. \label{eq:floper_e}
\end{align}
\end{subequations} 
The correlation between these fluctuations in the steady state is given by the symmetric covariance matrix $\boldsymbol{C}_{\infty}$, which has real entries in purely absorptive bistability. It contains nine independent elements following a reduction from fifteen steady-state correlations \cite{Carmichael1986}
\begin{widetext}
\begin{equation}\label{eq:covss}
\frac{\boldsymbol{C}_{\infty}}{N}=\begin{pmatrix}
\braket{\Delta \tilde{\bar{a}} \Delta \tilde{\bar{a}}}_{\rm ss} & \braket{\Delta \tilde{\bar{a}}^{\dagger} \Delta \tilde{\bar{a}}}_{\rm ss} & \braket{\Delta \tilde{\bar{a}} \Delta \tilde{\bar{J}}_{-}}_{\rm ss} & \braket{\Delta \tilde{\bar{a}} \Delta \tilde{\bar{J}}_{+}}_{\rm ss} & \braket{\Delta\tilde{\bar{a}} \Delta \bar{J}_z}_{\rm ss} \\
\cdots & \braket{\Delta \tilde{\bar{a}}^{\dagger} \Delta \tilde{\bar{a}}^{\dagger}}_{\rm ss} & \braket{\Delta \tilde{\bar{a}}^{\dagger} \Delta \tilde{\bar{J}}_{-}}_{\rm ss} & \braket{\Delta \tilde{\bar{a}}^{\dagger} \Delta \tilde{\bar{J}}_{+}}_{\rm ss} & \braket{\Delta \tilde{\bar{a}}^{\dagger}\Delta \bar{J}_{z}}_{\rm ss} \\
\cdots & \cdots & \braket{\Delta \tilde{\bar{J}}_{-} \Delta \tilde{\bar{J}}_{-}}_{\rm ss} & \braket{\Delta \tilde{\bar{J}}_{+} \Delta \tilde{\bar{J}}_{-}}_{\rm ss} & \braket{\Delta \bar{J}_{z}\Delta \tilde{\bar{J}}_{-}}_{\rm ss} \\
\cdots & \cdots & \cdots & \braket{\Delta \tilde{\bar{J}}_{+} \Delta \tilde{\bar{J}}_{+}}_{\rm ss} & \braket{\Delta \tilde{\bar{J}}_{+}\Delta \bar{J}_{z}}_{\rm ss} \\
\cdots & \cdots & \cdots & \cdots & \braket{\Delta \bar{J}_{z} \Delta \bar{J}_{z}}_{\rm ss}
\end{pmatrix}.
\end{equation}
\end{widetext}

Equipped with the formalism required for treating linearized quantum fluctuations about the steady state of absorptive bistability, we proceed to the calculation of the incoherent correlation spectrum and the variances of the quadrature phase polarization amplitudes. Our analysis follows closely the treatment developed in Secs. 15.2.3, 15.2.6 and 15.2.7 of \cite{QO2} for the transmitted light and forwards photon scattering in the weak-excitation limit, and is extended to the strong-excitation regime of absorptive optical bistability.

\section{Incoherent spectrum of atomic correlations}
\label{sec:corrincohspectr}

To calculate the incoherent spectrum of the collective atomic degree of freedom, we need to determine the first-order fluctuation correlation function. The quantity of interest is the third element in the column vector
\begin{equation}\label{eq:vectorCovM}
\begin{aligned}
&\boldsymbol{C}^{\nu_{*};i}_{\rm ss}(\tau) \equiv \begin{pmatrix}
C_{\rm ss}^{\tilde{\nu}_{*}\tilde{z}}(\tau) \\
C_{\rm ss}^{\tilde{\nu}_{*}\tilde{z}_{*}}(\tau)\\
C_{\rm ss}^{\tilde{\nu}_{*}\tilde{\nu}}(\tau)\\
C_{\rm ss}^{\tilde{\nu}_{*}\tilde{\nu}_{*}}(\tau)\\
C_{\rm ss}^{\tilde{\nu}_{*}\mu}(\tau)
\end{pmatrix}
=N \lim_{t\to\infty} \begin{pmatrix}
\braket{\Delta \tilde{\bar{J}}_{+}(t) \Delta\tilde{\bar{a}}(t+\tau)}\\
\braket{\Delta \tilde{\bar{J}}_{+}(t) \Delta\tilde{\bar{a}}^{\dagger}(t+\tau)}\\
\braket{\Delta \tilde{\bar{J}}_{+}(t) \Delta\tilde{\bar{J}}_{-}(t+\tau)}\\
\braket{\Delta \tilde{\bar{J}}_{+}(t) \Delta\tilde{\bar{J}}_{+}(t+\tau)}\\
\braket{\Delta \tilde{\bar{J}}_{+}(t) \Delta\bar{J}_{z}(t+\tau)}
\end{pmatrix},\\
&i=\tilde{z}, \tilde{z}_{*}, \tilde{\nu}, \tilde{\nu}_{*}, \mu,
\end{aligned}
\end{equation}
sequestered as the fourth row of the covariance matrix. From Eq. \eqref{eq:eomC}, this vector obeys the equation of motion
\begin{equation}\label{eq:vectorEoM}
\frac{d\boldsymbol{C}_{\rm ss}^{\nu_{*};i}}{d\tau}=\bar{\boldsymbol{J}}_{\rm ss}\boldsymbol{C}_{\rm ss}^{\nu_{*};i},
\end{equation}
with initial conditions (at $\tau=0$) determined by Eq. \eqref{eq:initialcondC}. Overlooking all spatial effects due to the different position of the atoms, as we did when we assumed an equal coupling strength to the intracavity field in ME \eqref{eq:BistSE}, the incoherent power spectrum for the collective atomic polarization as a system degree of freedom is defined as
\begin{equation}\label{eq:defspectrum}
\begin{aligned}
\overline{T}(y)&=\braket{\Delta{\tilde{\bar{J}}}_{+}\Delta\tilde{\bar{J}}_{-}}_{\rm ss}^{-1}\frac{1}{2\pi}\\
&\times \int_{-\infty}^{\infty} d\bar{\tau} \,e^{i[2(\omega-\omega_0)/\gamma]\bar{\tau}}\braket{\Delta\tilde{\bar{J}}_{+}(\bar{\tau})\Delta\tilde{\bar{J}}_{-}(0)}_{\rm ss}\\
&=\frac{1}{\pi C_{\rm ss}^{\tilde{\nu}_{*}\tilde{\nu}}(0)}{\rm Re}\{\bar{\mathcal{C}}_{\rm ss}^{\tilde{\nu}_{*}\tilde{\nu}}(\bar{s})\}_{\bar{s}=-i 2(\omega-\omega_0)/\gamma\equiv-iy},
\end{aligned}
\end{equation}
where $\mathcal{C}_{\rm ss}^{\tilde{\nu}_{*}\tilde{\nu}}(s)=(2/\gamma)\bar{\mathcal{C}}_{\rm ss}^{\tilde{\nu}_{*}\tilde{\nu}}(\bar{s})$ is the Laplace transform of $C_{\rm ss}^{\tilde{\nu}_{*}\tilde{\nu}}(\tau)$. The spectral distribution of Eq. \eqref{eq:defspectrum} is normalized to unity with respect to the dimensionless frequency $y\equiv 2(\omega-\omega_0)/\gamma$.

\subsection{Spectrum of cooperative atomic correlations in the weak-excitation limit}
\label{subsec:spewel}

In this section, we focus our attention on the regime of very weak intracavity excitation such that $X \ll X_{-}$. The Jacobian matrix of Eq. \eqref{eq:JacobFull} is approximated as
\begin{equation}\label{eq:Jacobapp}
\bar{\boldsymbol{J}}_{\rm ss}^{\rm w} \approx \frac{\gamma}{2} \begin{pmatrix}
-\xi & 0 & \xi 2C & 0 & 0 \\
0 & -\xi & 0 & \xi 2C & 0 \\
-1 & 0 & -1 & 0 & X \\
0 & -1 & 0 & -1 & X \\
X & X & -X & -X & -2
\end{pmatrix}.
\end{equation}
Using this form in Eqs. \eqref{eq:vectorEoM}, the equations of motion for the various correlation vector components read
\begin{subequations}\label{eq:system}
\begin{align}
\frac{d C_{\rm ss}^{\tilde{\nu}_{*}\tilde{z}}}{d\bar{\tau}}&=-\xi C_{\rm ss}^{\tilde{\nu}_{*}\tilde{z}} + \xi 2 C\, C_{\rm ss}^{\tilde{\nu}_{*}\tilde{\nu}}, \label{eq:system_a}\\
\frac{d C_{\rm ss}^{\tilde{\nu}_{*}\tilde{z}_{*}}}{d\bar{\tau}}&=-\xi C_{\rm ss}^{\tilde{\nu}_{*}\tilde{z}_{*}} + \xi 2 C\, C_{\rm ss}^{\tilde{\nu}_{*}\tilde{\nu}_{*}}, \label{eq:system_b}\\
\frac{d C_{\rm ss}^{\tilde{\nu}_{*}\tilde{\nu}}}{d\bar{\tau}}&=-C_{\rm ss}^{\tilde{\nu}_{*}\tilde{\nu}}-C_{\rm ss}^{\tilde{\nu}_{*}\tilde{z}} + X C_{\rm ss}^{\tilde{\nu}_{*}\mu}, \label{eq:system_c}\\
\frac{d C_{\rm ss}^{\tilde{\nu}_{*}\tilde{\nu}_{*}}}{d\bar{\tau}}&=-C_{\rm ss}^{\tilde{\nu}_{*}\tilde{\nu}_{*}}-C_{\rm ss}^{\tilde{\nu}_{*}\tilde{z}_{*}} + X C_{\rm ss}^{\tilde{\nu}_{*}\mu}, \label{eq:system_d}\\
\frac{d C_{\rm ss}^{\tilde{\nu}_{*}\mu}}{d\bar{\tau}}&=-2C_{\rm ss}^{\tilde{\nu}_{*}\mu} + X (C_{\rm ss}^{\tilde{\nu}_{*}\tilde{z}}+C_{\rm ss}^{\tilde{\nu}_{*}\tilde{z}_{*}}-C_{\rm ss}^{\tilde{\nu}_{*}\tilde{\nu}}-C_{\rm ss}^{\tilde{\nu}_{*}\tilde{\nu}_{*}}), \label{eq:system_e}
\end{align}
\end{subequations}
where $\bar{\tau}\equiv \gamma \tau/2$ is the dimensionless time. The initial conditions are produced by Eq. \eqref{eq:initialcondC} [they are also read from Sec. 15.2.3 of \cite{QO2} or Sec. III of \cite{Carmichael1986}] with $Y\approx (1+2C) X$ and $dY/dX \approx (1+2C)$:
\begin{subequations}\label{eq:initialcond}
\begin{align}
&C_{\rm ss}^{\tilde{\nu}_{*}\tilde{z}}(0)=X^4\frac{\xi 2C (2+\xi+2C)}{(1+2C)^2 (\xi+1)^2}, \label{eq:initialcond_a} \\
&C_{\rm ss}^{\tilde{\nu}_{*}\tilde{z}_{*}}(0)=-X^2 \frac{\xi 2C}{(1+2C)(\xi+1)}, \label{eq:initialcond_b} \\
&C_{\rm ss}^{\tilde{\nu}_{*}\tilde{\nu}}(0)=X^4 \frac{2C(2+\xi+2C)+(\xi+1)^2}{(1+2C)^2(\xi+1)^2}, \label{eq:initialcond_c} \\
&C_{\rm ss}^{\tilde{\nu}_{*}\tilde{\nu}_{*}}(0)=-X^2\frac{1+2C+2\xi}{(\xi+1)(1+2C)}, \label{eq:initialcond_d} \\
&C_{\rm ss}^{\tilde{\nu}_{*}\mu}(0)=X^3 \frac{2C+\xi+1}{(1+2C)(\xi+1)}.\label{eq:initialcond_e}
\end{align}
\end{subequations}
Keeping terms of the same order in $X$ on the right-hand sides of the equations comprising the system \eqref{eq:system} so as to match the left-hand side as read from the initial conditions \eqref{eq:initialcond} [which amounts to dropping $X C_{\rm ss}^{\tilde{\nu}_{*}\mu}$ from Eq. \eqref{eq:system_c} together with the terms $C_{\rm ss}^{\tilde{\nu}_{*}\tilde{z}}$, $C_{\rm ss}^{\tilde{\nu}_{*}\tilde{\nu}}$, from Eq. \eqref{eq:system_e}], the transformed equations are conveniently organized in the following three subsets:
\begin{subequations}\label{eq:LT}
\begin{align}
\begin{pmatrix}
\xi + \bar{s} & -\xi 2C \\
1 & 1+\bar{s}
\end{pmatrix} \begin{pmatrix}
\bar{\mathcal{C}}_{\rm ss}^{\tilde{\nu}_{*}\tilde{z}} \\
\bar{\mathcal{C}}_{\rm ss}^{\tilde{\nu}_{*}\tilde{\nu}}
\end{pmatrix}&= \begin{pmatrix}
C_{\rm ss}^{\tilde{\nu}_{*}\tilde{z}}(0) \\ C_{\rm ss}^{\tilde{\nu}_{*}\tilde{\nu}}(0)
\end{pmatrix}+ X \bar{\mathcal{C}}_{\rm ss}^{\tilde{\nu}_{*}\mu} \begin{pmatrix}
0 \\ 1
\end{pmatrix}, \label{eq:LT_a} \\
\begin{pmatrix}
\xi + \bar{s} & -\xi 2C \\
1 & 1+\bar{s}
\end{pmatrix} \begin{pmatrix}
\bar{\mathcal{C}}_{\rm ss}^{\tilde{\nu}_{*}\tilde{z}_{*}} \\
\bar{\mathcal{C}}_{\rm ss}^{\tilde{\nu}_{*}\tilde{\nu}_{*}}
\end{pmatrix}&=\begin{pmatrix}
C_{\rm ss}^{\tilde{\nu}_{*}\tilde{z}_{*}}(0) \\ C_{\rm ss}^{\tilde{\nu}_{*}\tilde{\nu}_{*}}(0)
\end{pmatrix}, \label{eq:LT_b} \\
(2+\bar{s}) \bar{\mathcal{C}}_{\rm ss}^{\tilde{\nu}_{*}\mu}&=C_{\rm ss}^{\tilde{\nu}_{*}\mu}(0) +X(\bar{\mathcal{C}}_{\rm ss}^{\tilde{\nu}_{*}\tilde{z}_{*}}-\bar{\mathcal{C}}_{\rm ss}^{\tilde{\nu}_{*}\tilde{\nu}_{*}}), \label{eq:LT_c}
\end{align}
\end{subequations}
in which we have introduced the scaled quantities:
\begin{equation}\label{eq:defLTv}
\bar{s}\equiv 2s/\gamma, \quad \mathcal{C}_{\rm ss}^{ij}(s)=\frac{2}{\gamma}\bar{\mathcal{C}}_{\rm ss}^{ij}(\bar{s}),
\end{equation}
where $\mathcal{C}_{\rm ss}^{ij}(\bar{s})$ is the Laplace transform of $C^{ij}_{\rm ss}(\tau)$, $i,j=\tilde{z}, \tilde{z}_{*}, \tilde{\nu}, \tilde{\nu}_{*}, \mu$. Using the inverse
\begin{equation}\label{eq:inv}
\begin{pmatrix}
\xi + \bar{s} & -\xi 2C \\
1 & 1+\bar{s}
\end{pmatrix}^{-1}=\frac{1}{(\xi +\bar{s})(1+\bar{s})+\xi 2C}\begin{pmatrix}
1 + \bar{s} & \xi 2C \\
-1 & \xi+\bar{s}
\end{pmatrix},
\end{equation}
we find from \eqref{eq:LT_b}
\begin{subequations}\label{eq:firsttwo}
\begin{align}
\bar{\mathcal{C}}_{\rm ss}^{\tilde{\nu}_{*}\tilde{z}_{*}}(\bar{s})&=-\frac{\xi 2C  X^2}{(\xi+1)(1+2C)} \frac{\bar{s}+\xi+2(C+1)}{(\xi +\bar{s})(1+\bar{s})+\xi 2C}, \label{eq:firsttwo_a} \\
\bar{\mathcal{C}}_{\rm ss}^{\tilde{\nu}_{*}\tilde{\nu}_{*}}(\bar{s})&=-\frac{X^2}{(\xi+1)(1+2C)} \frac{(1+\xi+2C)\bar{s}+\xi(\xi+1)}{(\xi +\bar{s})(1+\bar{s})+\xi 2C}. \label{eq:firsttwo_b}
\end{align}
\end{subequations}
Then, substituting into Eq. \eqref{eq:LT_c}, we obtain
\begin{equation}
\bar{\mathcal{C}}_{\rm ss}^{\tilde{\nu}_{*}\mu}(\bar{s})=X^3 \frac{(\xi+\bar{s})(\xi+1)+2C\bar{s}}{(1+2C)(\xi+1)[(\xi+\bar{s})(1+\bar{s})+\xi 2C]}.
\end{equation}
From Eq. \eqref{eq:LT_a}, we can solve for
\begin{equation}\label{eq:systemCL}
\begin{aligned}
& \begin{pmatrix}
\bar{\mathcal{C}}_{\rm ss}^{\tilde{\nu}_{*}\tilde{z}} \\
\bar{\mathcal{C}}_{\rm ss}^{\tilde{\nu}_{*}\tilde{\nu}}
\end{pmatrix}=\frac{1}{(\xi +\bar{s})(1+\bar{s})+\xi 2C}\\
& \times\begin{pmatrix}
1 + \bar{s} & \xi 2C \\
-1 & \xi+\bar{s}
\end{pmatrix} \left[\begin{pmatrix}
C_{\rm ss}^{\tilde{\nu}_{*}\tilde{z}}(0) \\ C_{\rm ss}^{\tilde{\nu}_{*}\tilde{\nu}}(0)
\end{pmatrix}+ X \bar{\mathcal{C}}_{\rm ss}^{\tilde{\nu}_{*}\mu} \begin{pmatrix}
0 \\ 1
\end{pmatrix}\right],
\end{aligned}
\end{equation}
whence the Laplace transform of the correlation function we seek is the bottom element of the vector on the left-hand side of Eq. \eqref{eq:systemCL}. We note the uniformity of all terms in the sum in the order $X^4$ since $X$ multiplies the correlation $C_{\rm ss}^{\tilde{\nu}_{*}\tilde{\nu}}(\bar{s})$, which is of order $X^3$. This guarantees the consistency of the small-$X$ expansion in the weak-excitation limit. The final expression reads 
\begin{equation}
\bar{\mathcal{C}}_{\rm ss}^{\tilde{\nu}_{*}\tilde{\nu}}(\bar{s})=\frac{(\xi+\bar{s})[C_{\rm ss}^{\tilde{\nu}_{*}\tilde{\nu}}(0)+X \bar{\mathcal{C}}_{\rm ss}^{\tilde{\nu}_{*}\mu}]-C_{\rm ss}^{\tilde{\nu}_{*}\tilde{z}}(0)}{(\xi +\bar{s})(1+\bar{s})+\xi 2C} ,
\end{equation}
or, separating the two denominators as the determinant of the $2\times 2$ matrix in Eq. \eqref{eq:systemCL} and its square, 
\begin{equation}\label{eq:CorrAA}
\begin{aligned}
&\bar{\mathcal{C}}_{\rm ss}^{\tilde{\nu}_{*}\tilde{\nu}}(\bar{s})=X^4 \frac{1}{(\xi+1)(1+2C)}\\
&\times\Bigg\{\frac{[2C(2+\xi+2C)+(\xi+1)^2]\bar{s}+\xi(\xi+1)^2}{(\xi+1)(1+2C)[(\xi +\bar{s})(1+\bar{s})+\xi 2C]} \\
&+\frac{(\xi+\bar{s})[(\xi+\bar{s})(\xi+1)+2C\bar{s}]}{[(\xi +\bar{s})(1+\bar{s})+\xi 2C]^2}\Bigg\}.
\end{aligned}
\end{equation}
Applying Eq. \eqref{eq:defspectrum} yields the {\it normalized incoherent spectrum of the collective atomic degree of freedom in the weak-excitation limit of absorptive bistability}:
\begin{equation}\label{eq:normspfinal}
\begin{aligned}
&\overline{T}(y)=\frac{1}{\pi}{\rm Re}\Bigg\{\frac{\xi(\xi+1)^2-iy A(\xi, C)}{A(\xi, C)[(\xi-iy)(1-iy)+\xi 2C]} \\
&+ \frac{(\xi+1)(1+2C)(\xi-iy)[(\xi-iy)(\xi+1)-2iCy]}{A(\xi, C)[(\xi-iy)(1-iy)+\xi 2C]^2}\Bigg\},
\end{aligned}
\end{equation}
where $A(\xi, C)\equiv 2C(2+\xi+2C)+(\xi+1)^2$. 

Using Eq. \eqref{eq:system_c} and the three initial conditions of Eqs. \eqref{eq:initialcond_a}, \eqref{eq:initialcond_c} and \eqref{eq:initialcond_e}, we can verify that 
\begin{equation}
\frac{d C_{\rm ss}^{\tilde{\nu}_{*}\tilde{\nu}}(\bar{\tau})}{d\bar{\tau}}\Bigg|_{\bar{\tau}=0}=0,
\end{equation}
from which we expect a $|\omega-\omega_0|^{-4}$ asymptotic behavior of the correlation spectrum in the weak-excitation limit (see Appendix B of \cite{Kochan1994} and Note 15.8 of \cite{QO2}). Next, we will focus on the three main limits arising in the operation of the bistable absorber at weak excitation, simplifying the expression of Eq. \eqref{eq:normspfinal}.

\subsubsection{Bad-cavity limit}
\label{subsubsec:bcl}

In the bad-cavity regime ($\xi \gg 1,2C$) we take the limit $\xi \to \infty$ in Eq. \eqref{eq:normspfinal}, yielding
\begin{equation}\label{eq:bdl}
\begin{aligned}
\overline{T}(y)&=\frac{1}{\pi} {\rm Re}\left[\frac{1}{1+2C-iy}+(1+2C)\left(\frac{1}{1+2C-iy}\right)^2\right]\\
&=\frac{1}{\pi} \frac{2(1+2C)^3}{[(1+2C)^2+y^2]^2}.
\end{aligned}
\end{equation}
Equation \eqref{eq:bdl} predicts a Lorentzian squared with the same form to that identified in \cite{RiceCarmichaelNonClassical} for free-space resonance fluorescence, where the presence of the square is attributed to squeezing. This distribution with a collectively enhanced linewidth, as depicted in Fig. \ref{fig:spectra}(a), is also attained for the forwards-scattered field in the bad cavity limit for weak excitation. Guided by the correspondence between the intracavity field and the atomic fluorescence, we will explore another two distinct limits of Eq. \eqref{eq:normspfinal}.

\subsubsection{Good-cavity limit}
\label{subsubsec:gcl}

In the good-cavity limit ($\xi \ll 1,2C$), we note that the distribution $T(y)$ is peaked about $y \sim \xi \ll 1$. With that observation, we obtain the approximate expression
\begin{equation}\label{eq:gql}
\begin{aligned}
\overline{T}(y)&=\frac{1}{\pi} {\rm Re}\Bigg\{\frac{\xi-i2y(\xi+2C)(C+1)}{2(\xi+2C)(C+1)[\xi(1+2C)-iy]}\\
& + \frac{(\xi+1)(1+2C)(\xi-iy)[\xi-iy(1+2C)]}{2(\xi+2C)(C+1)[\xi(1+2C)-iy]^2}\Bigg\}.
\end{aligned}
\end{equation}
The above approximation is valid only for $|y| \sim \xi$ and captures the effect of a {\it spectral hole} creation in the center of the distribution, as we can see in Fig. \ref{fig:spectra}(b). This effect is as well observed in the incoherent part of the optical spectrum for the transmitted light (see Fig. 5 of \cite{RiceCarmichaelNonClassical}) and is due to squeezing.

\subsubsection{Collective strong-coupling regime}
\label{subsubsec:collscr}

We will now move to a markedly different regime defined by the large product of the cooperativity parameter $2C$ and the dimensionless dissipation rate $\xi$. On factorizing the denominator of both terms in Eq. \eqref{eq:normspfinal} as 
\begin{equation}
(\xi-iy)(1-iy)+\xi 2C=(\bar{\lambda}_{+}+iy)(\bar{\lambda}_{-}+iy),
\end{equation}
with 
\begin{equation}\label{eq:eigenlambda}
\bar{\lambda}_{\pm}=-\frac{1}{2}(\xi+1)\pm i \sqrt{\xi 2C-\frac{1}{4}(\xi-1)^2},
\end{equation}
we observe that another limit arises in the many-atom strong coupling regime, namely for $(\xi+1) \ll 2\sqrt{\xi 2 C}$. We can then write [see Sec. 15.2.6. of \cite{QO2}]
\begin{equation}
\begin{aligned}
&(\xi-iy)(1-iy)+\xi 2C \approx P_{+}(y;\xi,C)P_{-}(y;\xi,C),
\end{aligned}
\end{equation} 
with
\begin{equation}
P_{\pm}(y;\xi,C)\equiv \frac{1}{2}(\xi+1)-i(y\pm\sqrt{\xi2C}).
\end{equation}
Consequently, we accept the following approximations for the first and second terms inside the real part of Eq. \eqref{eq:normspfinal}, respectively,
\begin{equation}
\begin{aligned}
\frac{\frac{1+\xi}{2}-iy}{(\xi-iy)(1-iy)+\xi 2C}& \approx \frac{1}{2}\left[\frac{1}{P_{+}(y;\xi,C)}+\frac{1}{P_{-}(y;\xi,C)}\right],
\end{aligned}
\end{equation}
and
\begin{equation}
\begin{aligned}
&\frac{\frac{(\xi+1)^2}{2}-2y^2-2iy(\xi+1)-\xi 4 C}{[(\xi-iy)(1-iy)+\xi 2C]^2}\\
& \approx \frac{1}{P^2_{+}(y;\xi,C)}+\frac{1}{P^2_{-}(y;\xi,C)}.
\end{aligned}
\end{equation}
Close to the resonances at $y=+\sqrt{\xi 2C}$ and $y=-\sqrt{\xi 2C}$ we replace $y$ in each nonresonant term by its value on resonance. With this assumption, the squared Lorentzians dominate, and after normalization we obtain
\begin{equation}\label{eq:spSCL}
\begin{aligned}
\overline{T}(y)&=\frac{1}{\pi}\Bigg\{\frac{\left[\frac{1}{2}(\xi+1)\right]^3}{\left\{ \left[\frac{1}{2}(\xi+1)\right]^2 + (y+\sqrt{\xi 2C})^2\right\}^2}\\
&+ \frac{\left[\frac{1}{2}(\xi+1)\right]^3}{\left\{ \left[\frac{1}{2}(\xi+1)\right]^2 + (y-\sqrt{\xi 2C})^2\right\}^2} \Bigg\}.
\end{aligned}
\end{equation} 
The above expression produces a spectrum with a vacuum Rabi doublet where every peak is a squared Lorentzian, exactly like what happens for the forwards scattered field due to squeezing of quantum fluctuations in the weak-excitation limit. An example of a Rabi doublet is depicted in Fig. \ref{fig:spectra}(c), where the two squared Lorentzians are centered at $y=\pm \sqrt{\xi 2C}=\pm 20$, while every feature at the center of the spectral distribution has disappeared [compare with Fig. 2(b) of \cite{Raizen1989} for the transmitted light]. The many-atom vacuum Rabi splitting arises as a distinct feature of collective strong driving in the weak-excitation limit in contrast to what we observe for a single driven radiatively damped atom where the occurrence of level splitting requires a strong drive (see Sec. V of \cite{Carmichael1986}).  

\subsection{Dynamical Stark shift along the upper branch}
\label{subsec:SEUB}

\begin{figure*}
\includegraphics[width=\textwidth]{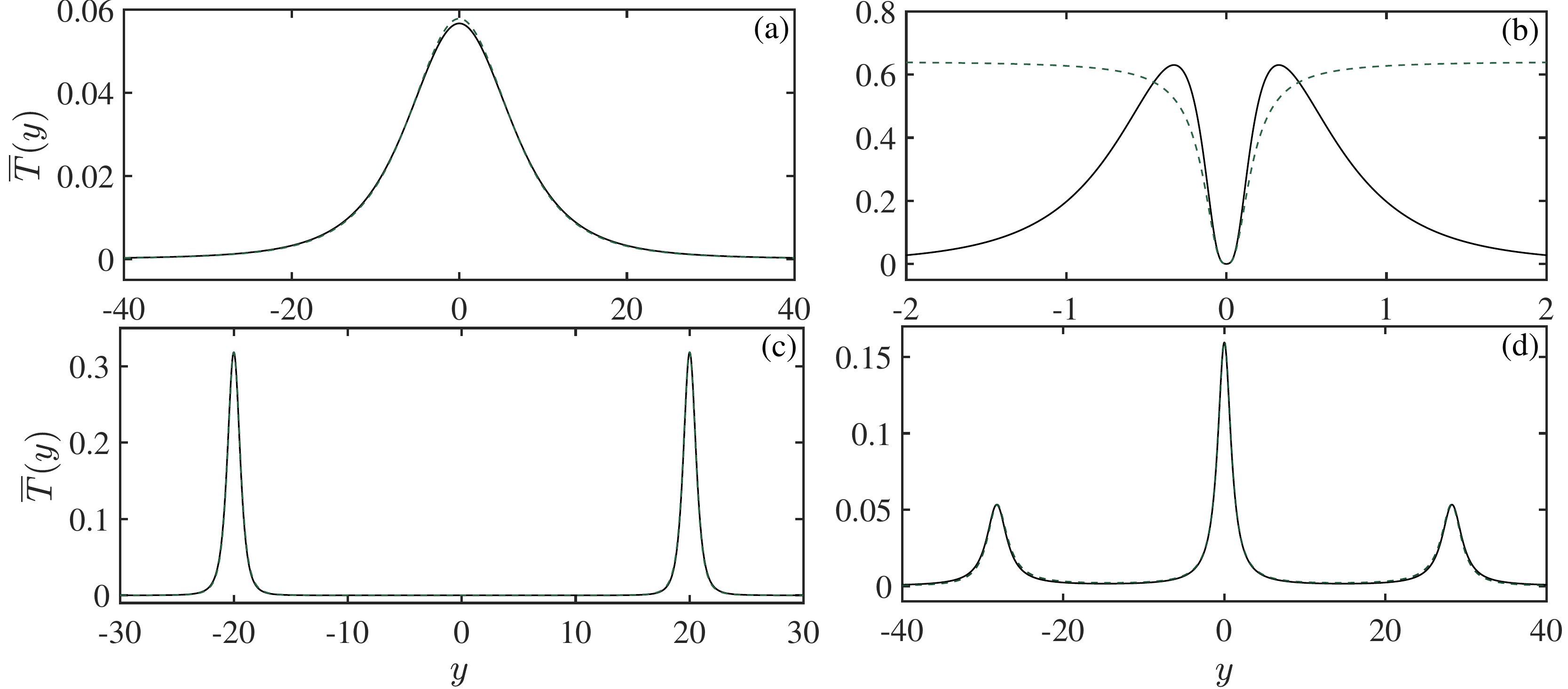}
\caption{{\it Normalized incoherent spectrum for the collective atomic polarization in the small-noise limit against the dimensionless frequency $y=2(\omega-\omega_0)/\gamma$.} {\bf (a)} The spectrum from the full analytical expression of Eq. \eqref{eq:normspfinal} in solid black line is plotted against the squared Lorentzian distribution of Eq. \eqref{eq:bdl} in dashed green in the bad-cavity limit for $C=5$, $\xi=500$. {\bf (b)} The spectrum from Eq. \eqref{eq:normspfinal} in solid black line is plotted against the spectral hole predicted by Eq. \eqref{eq:gql} in dashed green in the good-cavity limit for $C=5$, $\xi=0.01$. The two curves match only for $|y| \sim \xi$. {\bf (c)} The spectrum from Eq. \eqref{eq:normspfinal} in solid black line is plotted against the many-atom Rabi doublet predicted by Eq. \eqref{eq:spSCL} in dashed green in the good-cavity limit for $C=200$, $\xi=1$. {\bf (d)} The spectrum from Eq. \eqref{eq:TSELcoll} in solid black line is superimposed on the spectrum obtained from Eq. (4.7) of \cite{Agarwal1978} in dashed green for a scaled intracavity amplitude $X=20$. The two Stark-shifted peaks are centered at $y=\pm \sqrt{2} X \approx \pm 28.3$.}
\label{fig:spectra}
\end{figure*}

For the upper branch with $X \gg X_{+}$, the Jacobian matrix assumes the form
\begin{equation}\label{eq:JappSE}
\bar{\boldsymbol{J}}_{\rm ss}^{\rm s} \approx\frac{\gamma}{2} \begin{pmatrix}
-\xi & 0 & \xi 2C & 0 & 0 \\
0 & -\xi & 0 & \xi 2C & 0 \\
0 & 0 & -1 & 0 & X \\
0 & 0 & 0 & -1 & X \\
0 & 0 & -X & -X & -2
\end{pmatrix}.
\end{equation}
The equations of motion for the atomic variables are then decoupled, using the lower block of the Jacobian matrix,
\begin{equation}
\begin{pmatrix}
 -1 & 0 & X \\
 0 & -1 & X \\
-X & -X & -2
\end{pmatrix},
\end{equation}
which is the same as the matrix $M$ of Eq. (3.22) of \cite{Agarwal1978} for $Y \approx X$ and under the scaling of Eqs. \eqref{eq:scalingFP}. The resulting equations read
\begin{subequations}\label{eq:systemUB}
\begin{align}
\frac{d C_{\rm ss}^{\tilde{\nu}_{*}\tilde{\nu}}}{d\bar{\tau}}&=-C_{\rm ss}^{\tilde{\nu}_{*}\tilde{\nu}} + X C_{\rm ss}^{\tilde{\nu}_{*}\mu}, \label{eq:systemUB_a}\\
\frac{d C_{\rm ss}^{\tilde{\nu}_{*}\tilde{\nu}_{*}}}{d\bar{\tau}}&=-C_{\rm ss}^{\tilde{\nu}_{*}\tilde{\nu}_{*}} + X C_{\rm ss}^{\tilde{\nu}_{*}\mu}, \label{eq:systemUB_b}\\
\frac{d C_{\rm ss}^{\tilde{\nu}_{*}\mu}}{d\bar{\tau}}&=-2C_{\rm ss}^{\tilde{\nu}_{*}\mu}-X(C_{\rm ss}^{\tilde{\nu}_{*}\tilde{\nu}}+C_{\rm ss}^{\tilde{\nu}_{*}\tilde{\nu}_{*}}), \label{eq:systemUB_c}
\end{align}
\end{subequations}
while for the initial conditions we set $Y \approx X$, and $dY/dX \approx 1$  to obtain
\begin{subequations}\label{eq:initialcond_UB}
\begin{align}
&C_{\rm ss}^{\tilde{\nu}_{*}\tilde{\nu}}(0)=1, \label{eq:initialcondUB_a} \\
&C_{\rm ss}^{\tilde{\nu}_{*}\tilde{\nu}_{*}}(0)=0, \label{eq:initialcondUB_b} \\
&C_{\rm ss}^{\tilde{\nu}_{*}\mu}(0)=0.\label{eq:initialcondUB_c}
\end{align}
\end{subequations}
The transformed equations for the atomic correlations can be written in a simple system form, reminiscent of the optical Bloch equations for resonance fluorescence,
\begin{equation}
\begin{pmatrix}
1+\bar{s} & 0 & -X \\
0 & 1+\bar{s} &  -X \\
X & X & 2+\bar{s}
\end{pmatrix} \begin{pmatrix}
\bar{\mathcal{C}}_{\rm ss}^{\tilde{\nu}_{*}\tilde{\nu}} \\
\bar{\mathcal{C}}_{\rm ss}^{\tilde{\nu}_{*}\tilde{\nu}_{*}}\\
\bar{\mathcal{C}}_{\rm ss}^{\tilde{\nu}_{*}\mu}
\end{pmatrix}=\begin{pmatrix}
C_{\rm ss}^{\tilde{\nu}_{*}\tilde{\nu}}(0) \\
C_{\rm ss}^{\tilde{\nu}_{*}\tilde{\nu}_{*}}(0)\\
C_{\rm ss}^{\tilde{\nu}_{*}\mu}(0)
\end{pmatrix}.
\end{equation}
Following the usual prescription of Eq. \eqref{eq:defspectrum}, after identifying $\bar{\mathcal{C}}_{\rm ss}^{\tilde{\nu}_{*}\tilde{\nu}}(\bar{s})$, the {\it correlation spectrum in the upper branch of the steady-state bistability curve far away from the turning point} is given by the expression
\begin{equation}\label{eq:TSELcoll}
\overline{T}(y)=\frac{1}{\pi}{\rm Re}\left\{\frac{X^2+(1-iy)(2-iy)}{(1-iy)[2X^2 + (1-iy)(2-iy)]}\right\},
\end{equation}
with $y\equiv 2(\omega-\omega_0)/\gamma$. The spectral distribution predicted by Eq. \eqref{eq:TSELcoll} is plotted in Fig. \ref{fig:spectra}(d) and is compared to the `limit of vanishing atomic density' when the cavity field is adiabatically eliminated [see Eq. (4.7) of \cite{Agarwal1978}]. The two spectra are practically indistinguishable from each other; the two Stark-shifted peaks are centered at $y=\pm \sqrt{2} X$ [see Eq. \eqref{eq:g2SC} and the eigenvalues of the linearized Maxwell-Bloch equations plotted in Fig. 1 of \cite{Raizen1987}] and compare with the transmitted spectrum in Fig. 2(b) of \cite{Carmichael1983}; compare also with the fluorescent spectra in Fig. 4 of \cite{Gladush2011} for the upper branch of dispersive bistability where the slope of the input-output curve visibly deviates from unity]. We find a ratio $3:1$ between the height of the central peak and that of the sidebands, as in \cite{Bonifacio1978PhysRevLett} for the bad-cavity limit [see also Fig. 3(d) therein]. We need to emphasize, however, that this result does not pertain to the fluorescent spectrum itself since unlike-atom correlations do not add up constructively to output the correlation function of the system atomic degrees of freedom (see Sec. \ref{sec:accessatem}). We also note that the distribution of Eq. \eqref{eq:TSELcoll} is independent of $\xi$ and $C$ and thus applicable to all limits we have discussed in Sec. \ref{subsec:spewel}, marking the lack of co-operation between the emitters coupled to the cavity mode \textemdash{a} justification for calling the upper branch the {\it independent atom branch}. Well separated sidebands are also predicted for the incoherent spectrum of the transmitted light in the mean-field analysis of \cite{Bonifacio1978} for $C \gg 1$ and $Y>C$, within the framework of a discontinuous band formation from a centrally symmetric narrow spectral distribution. For the fluorescent spectrum calculated in the bad-cavity limit, however, the Stark triplet is present along the entire upper branch while no narrowing of the central peak is predicted, following the suppression of collective effects \cite{Carmichael1981FS}.  

To determine the autocorrelation function of the cavity-field amplitude and the spectral distribution of the forwards-scattered field, one follows the same procedure described in Sec. \ref{subsec:spewel} but instead seeking $\mathcal{C}_{\rm ss}^{\tilde{z}_{*}\tilde{z}}(s)=(2/\gamma)\bar{\mathcal{C}}_{\rm ss}^{\tilde{z}_{*}\tilde{z}}(\bar{s})$, where $\mathcal{C}_{\rm ss}^{\tilde{z}_{*}\tilde{z}}(s)$ is the Laplace transform of $C_{\rm ss}^{\tilde{z}_{*}\tilde{z}}(\tau)$, extracted from the second row of the covariance matrix defined in Eq. \eqref{eq:defCovMatrix}. The steps required for the calculation are described in Sec. 15.2.6 of \cite{QO2}. For the dimensionless spectrum, which is normalized to unity, we then obtain
\begin{equation}\label{eq:defspectrumFD}
\begin{aligned}
\overline{T}_{\rm F}(y)&=\braket{\Delta{\tilde{\bar{a}}}^{\dagger}\Delta\tilde{\bar{a}}}_{\rm ss}^{-1}\frac{1}{2\pi}\\
&\times\int_{-\infty}^{\infty} d\bar{\tau} \,e^{i[2(\omega-\omega_0)/\gamma]\bar{\tau}}\braket{\Delta\tilde{\bar{a}}^{\dagger}(\bar{\tau})\Delta\tilde{\bar{a}}(0)}_{\rm ss}\\
&=\frac{1}{\pi C_{\rm ss}^{\tilde{z}_{*}\tilde{z}}(0)}{\rm Re}\{\bar{\mathcal{C}}_{\rm ss}^{\tilde{z}_{*}\tilde{z}}(\bar{s})\}_{\bar{s}=-2i(\omega-\omega_0)/\gamma\equiv-iy}.
\end{aligned}
\end{equation}
The initial conditions in the strong-excitation limit read \footnote{Care must be taken with the equations giving $C_{\rm ss}^{\tilde{z}_{*}\tilde{z}_{*}}(0)$ and $C_{\rm ss}^{\tilde{z}_{*}\tilde{\nu}_{*}}(0)$ containing the term $X^2(1-X/Y) \approx 2C$. These correlators, however, are not relevant to the calculation of the collective atomic correlation spectrum for $X \gg X_{+}$, since the correlations containing only atomic fluctuations are decoupled from the intracavity-field fluctuations.}
\begin{subequations}\label{eq:initialcondSE}
\begin{align}
&C_{\rm ss}^{\tilde{z}_{*}\tilde{z}}(0)=4C^2 \frac{\xi}{\xi+1} K(X, \xi), \label{eq:initialcondSE_a} \\
&C_{\rm ss}^{\tilde{z}_{*}\tilde{z}_{*}}(0)=4C^2 \frac{\xi}{\xi+1}[K(X, \xi)-1], \label{eq:initialcondSE_b} \\
&C_{\rm ss}^{\tilde{z}_{*}\tilde{\nu}}(0)=2C\frac{\xi}{\xi+1}K(X, \xi), \label{eq:initialcondSE_c} \\
&C_{\rm ss}^{\tilde{z}_{*}\tilde{\nu}_{*}}(0)=2C\frac{\xi}{\xi+1}[K(X, \xi)-1], \label{eq:initialcondSE_d} \\
&C_{\rm ss}^{\tilde{z}_{*}\mu}(0)\sim 1/X \approx 0,\label{eq:initialcondSE_e}
\end{align}
\end{subequations}
where $K(X,\xi)\equiv[X^2+(\xi+1)(\xi+3)]/[2X^2 + \xi(\xi+3)]$ is a saturation factor placing $X$ and $\xi$ on a similar footing. Based on the Jacobian matrix of Eq. \eqref{eq:JappSE}, we write
\begin{equation}\label{eq:CzzSE}
(\bar{s}+\xi)\bar{\mathcal{C}}_{\rm ss}^{\tilde{z}_{*}\tilde{z}}(\bar{s})=C_{\rm ss}^{\tilde{z}_{*}\tilde{z}}(0)+2\xi C \bar{\mathcal{C}}_{\rm ss}^{\tilde{z}_{*}\tilde{\nu}}(\bar{s}),
\end{equation}
with
\begin{equation}
\bar{\mathcal{C}}_{\rm ss}^{\tilde{z}_{*}\tilde{\nu}}(\bar{s})=\frac{[X^2+(\bar{s}+1)(\bar{s}+2)]C_{\rm ss}^{\tilde{z}_{*}\tilde{\nu}}(0)-X^2 C_{\rm ss}^{\tilde{z}_{*}\tilde{\nu}_{*}}(0)}{(\bar{s}+1)[2X^2+(\bar{s}+1)(\bar{s}+2)]}.
\end{equation}
For $X \gg \xi$, giving $K(X,\xi) \approx 1/2$, we obtain the simple expression after solving for $\bar{\mathcal{C}}_{\rm ss}^{\tilde{z}_{*}\tilde{z}}(\bar{s})$,
\begin{equation}\label{eq:SpFSE}
\overline{T}_{\rm F}(y)=\frac{1}{\pi}{\rm Re}\left[\frac{1+\xi -iy}{(\xi-iy)(1-iy)}\right],
\end{equation}
which is a Lorentzian distribution independent of the intracavity excitation and of width $\sim \xi$. On the other hand,  in the bad-cavity limit defined as $\xi \gg X$, we obtain $K(X,\xi) \approx 1$, whence (omitting terms with prefactors of order $\xi^{-1}$)
\begin{equation}\label{eq:BCUB}
\begin{aligned}
 &\overline{T}_{\rm F}(y)\approx \frac{1}{\pi}{\rm Re}\Bigg\{ \frac{2}{\xi-iy}\\
 &+ \frac{\xi[2X^2+2(1-iy)(2-iy)]}{[2X^2+(1-iy)(2-iy)](1-iy)(\xi-iy)} \Bigg\},
 \end{aligned}
\end{equation}
restoring the familiar Stark-triplet spectrum with the sidebands centered at $y=\pm \sqrt{2}X$, and a central to sideband peak height ratio approximately equal to $3:1$. 

\section{Accessing the collective atomic emission}
\label{sec:accessatem}

Let us now discuss the physical accessibility of the system collective degrees of freedom, as dictated by the ME \eqref{eq:MEbist}. We will first look at the field scattered into the infinitesimal solid angle $d\Omega$ in the direction $\hat{r}\equiv(\theta,\phi)=r^{-1}\boldsymbol{r}$ (where $\boldsymbol{r}$ is the field-observation position vector from the center of coordinates), expressed in photon flux units, which reads \cite{QO2}
\begin{equation}
\hat{\mathcal{E}}_{\rm s}=\sqrt{\frac{3d\Omega}{8\pi}}\sin\theta \sqrt{\gamma}\sum_{j=1}^{N}e^{-i\ k \hat{r}\cdot \boldsymbol{r}_j}\,\sigma_{j-}(t^{\prime}),
\end{equation}
where $k \equiv \omega_0/c$ is the wavenumber and $t^{\prime}\equiv t-r/c$ is the retarded time with respect to which all steady-state averages are considered. Hence, the incoherent scattered flux can be calculated as
\begin{equation}
\braket{\Delta\hat{\tilde{\mathcal{E}}}^{\dagger}_{\rm sc}\Delta\hat{\tilde{\mathcal{E}}}_{\rm sc}}_{\rm ss}= N\frac{3\gamma d\Omega}{8\pi}\sin^2\theta \,\, \mathcal{F}(\hat{r}, \xi, C, X),
\end{equation}
with
\begin{equation}\label{eq:ScatteredInc}
\begin{aligned}
&\mathcal{F}(\hat{r}, \xi, C, X) \equiv \frac{1}{N}\sum_{j,k=1}^{N}e^{-ik_0 \hat{r}\cdot (\boldsymbol{r}_j-\boldsymbol{r}_k)}\,\braket{\Delta\tilde{\sigma}_{k+} \Delta\tilde{\sigma}_{j-}}_{\rm ss}\\
&=\braket{\Delta \tilde{\sigma}_{j+} \Delta\tilde{\sigma}_{j-}}_{\rm ss} + \frac{1}{N} \sum_{\substack{j,k=1 \\ j\neq k}}^{N}e^{-ik_0 \hat{r}\cdot (\boldsymbol{r}_j-\boldsymbol{r}_k)}\,\braket{\Delta\tilde{\sigma}_{k+} \Delta\tilde{\sigma}_{j-}}_{\rm ss} \\
&=\braket{\Delta \tilde{\sigma}_{+} \Delta\tilde{\sigma}_{-}}_{\rm like}^{\rm ss} + \braket{\Delta \tilde{\sigma}_{+} \Delta\tilde{\sigma}_{-}}_{\rm unlike}^{\rm ss} \frac{1}{N}\sum_{\substack{j,k=1 \\ j\neq k}}^{N}e^{-i k_0 \hat{r}\cdot (\boldsymbol{r}_j-\boldsymbol{r}_k)},
\end{aligned}
\end{equation}
where $\braket{\Delta \tilde{\sigma}_{+} \Delta\tilde{\sigma}_{-}}_{\rm like}^{\rm ss} \equiv \braket{\tilde{\sigma}_{j+}\tilde{\sigma}_{j-}}-\braket{\tilde{\sigma}_{j+}} \braket{\tilde{\sigma}_{j-}}$ for any atom $j$ in the collection, while  $\braket{\Delta \tilde{\sigma}_{+} \Delta\tilde{\sigma}_{-}}_{\rm unlike}^{\rm ss} \equiv \braket{\tilde{\sigma}_{j+}\tilde{\sigma}_{k-}}-\braket{\tilde{\sigma}_{j+}} \braket{\tilde{\sigma}_{k-}}$ for any two atoms $j$ and $k$ with $j \neq k$ [see Sec. 15.2.4 of \cite{QO2}]. Unlike-atom correlations are of order $N^{-1}$ individually, yet, as they are $N(N-1)$ in number, they contribute overall at the same order as the like-atom correlations to the incoherent scattered intensity \cite{QO2}. Since the atoms couple to the resonant cavity mode with the same strength and radiatively decay at the same rate $\gamma$, they are interchangeable in any operator average. At this point, the usual assumption \cite{Carmichael1981FS} 
\begin{equation}\label{eq:phaseassumpt}
\sum_{\substack{j,k=1 \\ j\neq k}}^{N}e^{-i k_0 \hat{r}\cdot (\boldsymbol{r}_j-\boldsymbol{r}_k)} \approx 0,
\end{equation}
in Eq. \eqref{eq:ScatteredInc}, leads to the scattered-field flux assuming the form of a summed single-atom output [see also Eqs. (35)-(37) of \cite{Carmichael1981FS} for the fluorescent spectrum],
\begin{equation}\label{eq:ScatteredIncFinal}
\braket{\Delta\hat{\tilde{\mathcal{E}}}^{\dagger}_{\rm sc}\Delta\hat{\tilde{\mathcal{E}}}_{\rm sc}}_{\rm ss}= N\frac{3\gamma d\Omega}{8\pi}\sin^2\theta \frac{1}{2}\left(\frac{X^2}{1+X^2}\right)^2,
\end{equation}
consistent with the corresponding Lindblad term in the ME of optical bistability,
\begin{equation}
\frac{\gamma}{2}\left(\sum_{j=1}^{N}2\sigma_{j-}\rho \sigma_{j+}-\frac{1}{2}J_{z}\rho - \frac{1}{2}\rho J_{z} -N\rho\right).
\end{equation}
The above correspondence is also consistent with the distinguishable nature of the scatterers. Atoms remain far apart compared with the wavelength $\lambda=2\pi c /\omega_0$, allowing in principle separate imaging of the fluorescence emitted from each one of them individually. We also observe that in the weak-excitation limit, Eq. \eqref{eq:ScatteredIncFinal} corresponds to the free-space resonance fluorescence of $N$ emitters, coherently excited with a reduced amplitude $Y/(1+2C)$ \textemdash{the} mean intracavity amplitude. The scattered flux is proportional to $X^4$, in agreement with the correlation of Eq. \eqref{eq:CorrAA}. We arrive at the same conclusion if we recast the scattered flux as
\begin{equation}\label{eq:ScatteredIncFinalYX}
\braket{\Delta\hat{\tilde{\mathcal{E}}}^{\dagger}_{\rm sc}\Delta\hat{\tilde{\mathcal{E}}}_{\rm sc}}_{\rm ss}= N\frac{3\gamma d\Omega}{8\pi}\sin^2\theta \frac{1}{2}X^2\left(\frac{Y-X}{2C}\right)^2,
\end{equation}
with $Y \approx (1+2C)X$ along the initial segment of the lower branch. In other words, the cooperativity parameter $2C$ cancels out explicitly and the photon flux depends only on the weak intracavity excitation,
\begin{equation}
 \braket{\Delta\hat{\tilde{\mathcal{E}}}^{\dagger}_{\rm sc}\Delta\hat{\tilde{\mathcal{E}}}_{\rm sc}}_{\rm ss}^{\rm weak \, excitation}\approx \frac{3 d\Omega}{8\pi}\sin^2\theta R_{\gamma} X^2,
\end{equation}
where $R_{\gamma} \equiv \gamma N X^2/2$ is the total spontaneous emission rate in that limit \cite{QO2}. Finally, we note that in the good-cavity limit ($\xi \to 0$) we can obviate the assumption of Eq. \eqref{eq:phaseassumpt}, since $\braket{\Delta \tilde{\sigma}_{+} \Delta\tilde{\sigma}_{-}}_{\rm unlike}^{\rm ss} \to 0$ to dominant order in $N^{-1}$ (see Sec. 15.2.4 of \cite{QO2}). 
\begin{figure}
\centering
\includegraphics[width=0.45\textwidth]{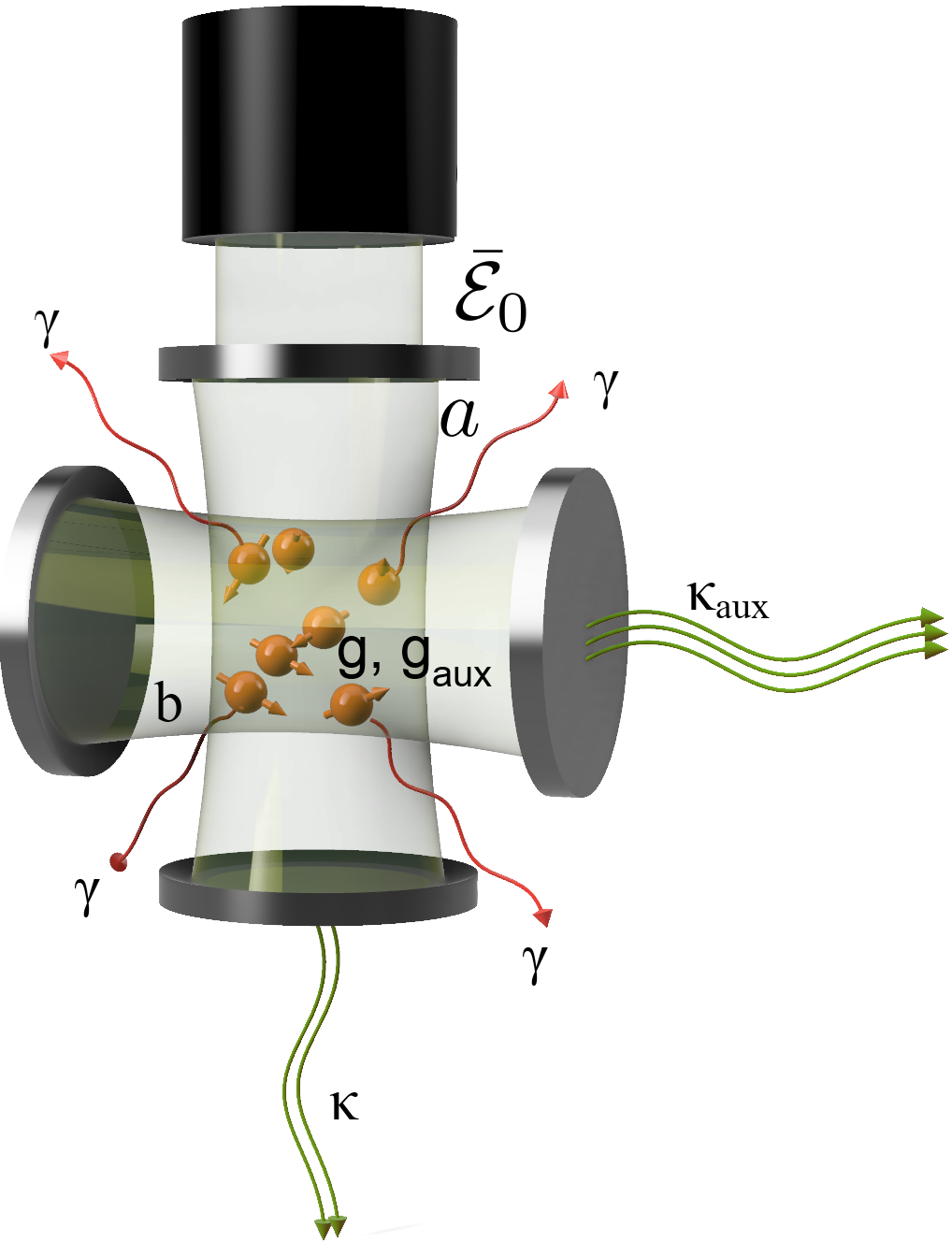}
\caption{{\it Schematic of the experimental setup proposed to extract collective emission.} The bistable-system cavity mode $a$, coupled to the atomic ensemble with strength $g$, is coherently driven with amplitude $\bar{\mathcal{E}}_0$. The interatomic distance is much larger than the resonant wavelength. The distinguishable scatterers are radiatively coupled to the reservoir modes with rate $\gamma$, and the photon loss rate is $2\kappa$. An auxiliary cavity mode $b$ is weakly coupled to the atomic ensemble with $g_{\rm aux} \ll g$. Operating in the bad-cavity limit ($\kappa_{\rm aux} \gg g_{\rm aux}$) allows for the adiabatic elimination of the auxiliary cavity field, giving access to the collective atomic polarization $J_{-}$.}
\label{fig:setup}
\end{figure}

How can one then access the atomic polarization, which is a internal system degree of freedom, translated into the output of a collective atomic emission channel? The answer lies in involving an auxiliary cavity mode supported by a low-Q cavity and coupled to the same atomic ensemble, as proposed by \cite{CarmichaelPC} and depicted in Fig. \ref{fig:setup}. If the Hamiltonian $H_{\rm aux}=\hbar \omega_0 b^{\dagger}b -i\hbar g_{\rm aux} (b^{\dagger}J_{-}-b J_{+})$ is added to the coherent part of the evolution in the ME \eqref{eq:MEbist}, with $g_{\rm aux} \ll g$, alongside the dissipation term $\kappa_{\rm aux}(2b\rho b^{\dagger}-b^{\dagger}b\rho-\rho b^{\dagger}b)$, with $\kappa_{\rm aux} \gg g_{\rm aux}$, then one may adiabatically eliminate the auxiliary cavity-field to produce (in the Heisenberg picture)
\begin{equation}
\tilde{b}(t) \approx \frac{g_{\rm aux}}{\kappa_{\rm aux}} \tilde{J}_{-}(t) + \kappa_{\rm aux}^{-1} \hat{\xi}(t),
\end{equation}
where the last term is due to the vacuum-field contribution with $\braket{\hat{\xi}(t)}=0$. It follows that the auxiliary-field fluctuation correlation function
\begin{equation}\label{eq:bcorr}
\kappa_{\rm aux} \braket{\Delta\tilde{b}^{\dagger}(0)\Delta \tilde{b}(\tau)}_{\rm ss} \approx \frac{g_{\rm aux}^2}{\kappa_{\rm aux}}\braket{\Delta\tilde{J}_{+}(0)\Delta \tilde{J}_{-}(\tau)}_{\rm ss},
\end{equation}
corresponding to a very weak output photon flux, is directly proportional to the collective atomic correlation function. Since $(g_{\rm aux}^2/\kappa_{\rm aux}) \ll g$, the effect of coupling the auxiliary mode to the collective atomic degrees of freedom is negligible even in the weak-excitation limit. Normalizing Eq. \eqref{eq:bcorr} by the incoherent transmitted photon flux $\kappa_{\rm aux}\braket{\Delta\tilde{b}^{\dagger}\Delta \tilde{b}}_{\rm ss}$ and taking the Fourier transform yields the collective correlation spectrum of Eq. \eqref{eq:defspectrum} which is hereby translated into the incoherent fluctuation spectrum of a single collective mode. 

\section{Squeezing of steady-state atomic fluctuations}
\label{sec:sqafl}

Squeezing is intimately tied to optical bistability and can already be deduced from the linear theory of quantum fluctuations. The analysis of \cite{Collett1985}, for example, demonstrated that perfect squeezing is in principle possible at the turning points of the steady-state input-output curve, while a more detailed analysis for the squeezing spectrum of dispersive bistability was carried out in \cite{Reid1985, Castelli1988}. To explicitly demonstrate the presence of squeezing, using the scaling relations \eqref{eq:ScalingOp_c} and \eqref{eq:ScalingOp_d}, we define the quadrature operators
\begin{equation}\label{eq:quadroper}
\begin{aligned}
&\Delta \tilde{J}_0 \equiv \frac{1}{\sqrt{2N}}(\Delta \tilde{J}_{-}\, e^{-i\phi_0} + \Delta \tilde{J}_{+}\, e^{i\phi_0}),\\
& \Delta \tilde{J}_{\pi/2} \equiv \frac{1}{\sqrt{2N}}[e^{-i(\phi_0+\pi/2)} \Delta \tilde{J}_{-} + e^{i(\phi_0+\pi/2)}\Delta \tilde{J}_{+}],
\end{aligned}
\end{equation}
with steady-state variances
\begin{equation}\label{eq:var1}
\begin{aligned}
(\Delta \tilde{J}_0)^2_{\rm ss}&=\frac{1}{2} N[\braket{\Delta \tilde{\bar{J}}_{+} \Delta \tilde{\bar{J}}_{-}}_{\rm ss}-\braket{\Delta \tilde{\bar{J}}_{+} \Delta \tilde{\bar{J}}_{+}}_{\rm ss}] - \frac{1}{4}\braket{\bar{J}_z}_{\rm ss}\\
&=\frac{1}{2}[C_{\rm ss}^{\tilde{\nu}_{*}\tilde{\nu}}(0)-C_{\rm ss}^{\tilde{\nu}_{*}\tilde{\nu}_{*}}(0)] - \frac{1}{4}\braket{\bar{J}_z}_{\rm ss},
\end{aligned}
\end{equation}
and
\begin{equation}\label{eq:var2}
\begin{aligned}
(\Delta \tilde{J}_{\pi/2})^2_{\rm ss}&=\frac{1}{2} N[\braket{\Delta \tilde{\bar{J}}_{+} \Delta \tilde{\bar{J}}_{-}}_{\rm ss}+\braket{\Delta \tilde{\bar{J}}_{+} \Delta \tilde{\bar{J}}_{+}}_{\rm ss}] - \frac{1}{4}\braket{\bar{J}_z}_{\rm ss}\\
&=\frac{1}{2}[C_{\rm ss}^{\tilde{\nu}_{*}\tilde{\nu}}(0)+C_{\rm ss}^{\tilde{\nu}_{*}\tilde{\nu}_{*}}(0)] - \frac{1}{4}\braket{\bar{J}_z}_{\rm ss},
\end{aligned}
\end{equation}
respectively, where the correlation functions are read from Eqs. \eqref{eq:initialcond_c} and \eqref{eq:initialcond_d} in the weak-excitation limit, with $\braket{\bar{J}_z}_{\rm ss} \approx -1$. Squeezing of fluctuations occurs when $C_{\rm ss}^{\tilde{\nu}_{*}\tilde{\nu}}(0)+C_{\rm ss}^{\tilde{\nu}_{*}\tilde{\nu}_{*}}(0)<0$, which is indeed the case for small intracavity amplitudes, since the ratio of the quadrature amplitude fluctuations over the intensity fluctuations is negative, while $(|\braket{\Delta \tilde{\bar{J}}_{+} \Delta \tilde{\bar{J}}_{+}}_{\rm ss}|/\braket{\Delta \tilde{\bar{J}}_{+} \Delta \tilde{\bar{J}}_{-}}_{\rm ss})$ diverges as $\sim 1/X^2$, violating the classical bound of unity (see also Sec. 15.2.3 of \cite{QO2}). In the bad-cavity limit ($\xi \gg 2C$) this divergence is further accentuated by the atomic cooperativity. We note that along much of the lower branch, $|\braket{\Delta \tilde{\bar{J}}_{+} \Delta \tilde{\bar{J}}_{+}}_{\rm ss}|>\braket{\Delta \tilde{\bar{J}}_{+} \Delta \tilde{\bar{J}}_{-}}_{\rm ss}$, which is not possible for a classical stochastic field. The negative sign for the normal-ordered variances of the squeezed forwards-scattered field may be picked by means of a conditional homodyne detection scheme proposed in \cite{Carmichael2000}. Squeezing for many atoms coupled to a cavity mode was experimentally observed in \cite{Raizen1987}, where the spectral density of fluctuations was measured at a particular frequency within the squeezing spectrum in a balanced homodyne detection scheme.

Such a divergence for the aforementioned ratio of fluctuations, however, does not occur in the strong-excitation limit. As one could anticipate from the Lorentzian spectrum of Eq. \eqref{eq:SpFSE}, when computing
\begin{equation}
\frac{\left|\braket{\Delta \tilde{\overline{a}}\Delta \tilde{\overline{a}}}_{\rm ss}\right|}{\braket{\Delta \tilde{\overline{a}}^{\dagger}\Delta \tilde{\overline{a}}}_{\rm ss}}=\frac{|C_{\rm ss}^{\tilde{z}\tilde{z}}(0)|}{C_{\rm ss}^{\tilde{z}_{*}\tilde{z}}(0)}=\frac{|C_{\rm ss}^{\tilde{z}_{*}\tilde{z}_{*}}(0)|}{C_{\rm ss}^{\tilde{z}_{*}\tilde{z}}(0)}=1,
\end{equation}
we obtain the expected upper bound for classical light with no presence of squeezing [compare also with the asymptotic values of $N\braket{\Delta \tilde{\overline{a}}^{\dagger}\Delta \tilde{\overline{a}}}_{\rm ss}$ and $N\braket{\Delta \tilde{\overline{a}}\Delta \tilde{\overline{a}}}_{\rm ss}$ in Figs. 15.1 (d)-(f) of \cite{QO2}, which are in agreement with Eqs. \eqref{eq:initialcondSE_a} and \eqref{eq:initialcondSE_b}]. The same conclusion is drawn for the collective atomic correlations along the independent branch. We also recall that in the single-atom free-space resonance fluorescence, phase information is destroyed in the strong-field limit and intensity correlations win over self-homodyning (see \cite{RiceCarmichaelIEEE} and Sec. 2.3.6 of \cite{QO1}). 

\section{Second-order coherence of atomic correlations}
\label{sec:2ndorderC}

As a further application of the linear theory of quantum fluctuations \textemdash{valid} provided that $N^{-1}$ remains the smallest parameter in the system \textemdash{we} will discuss the second-order coherence properties of the collective atomic polarization in the two extreme regions of the bistability state equations, where once again analytical expressions can be found. The normalized intensity correlation function is defined by the expression 
\begin{equation}\label{eq:defg2}
g^{(2)}(\tau)=\frac{\braket{\tilde{\bar{J}}_{+}(0)\tilde{\bar{J}}_{+}(\tau)\tilde{\bar{J}}_{-}(\tau)\tilde{\bar{J}}_{-}(0)}_{\rm ss}}{\left(\braket{\tilde{\bar{J}}_{+}\tilde{\bar{J}}_{-}}_{\rm ss}\right)^2},
\end{equation}
where 
\begin{equation*}
\begin{aligned}
&\braket{\tilde{\bar{J}}_{+}(0)\tilde{\bar{J}}_{+}(\tau)\tilde{\bar{J}}_{-}(\tau)\tilde{\bar{J}}_{-}(0)}_{\rm ss}\\
& \equiv \lim_{t \to \infty}\braket{\tilde{\bar{J}}_{+}(t)\tilde{\bar{J}}_{+}(t+\tau)\tilde{\bar{J}}_{-}(t+\tau)\tilde{\bar{J}}_{-}(t)}.
\end{aligned}
\end{equation*}
Expanding the collective atomic operators as sums of a mean and a fluctuation component, we can write
\begin{equation}
\braket{\tilde{\bar{J}}_{+}\tilde{\bar{J}}_{-}}_{\rm ss}=\braket{\tilde{\bar{J}}_{+}}_{\rm ss}\braket{\tilde{\bar{J}}_{-}}_{\rm ss} + \braket{\Delta\tilde{\bar{J}}_{+}\Delta\tilde{\bar{J}}_{-}}_{\rm ss}.
\end{equation}
Expanding the numerator of the correlation function yields
\begin{equation}\label{eq:numg2}
\begin{aligned}
&\braket{\tilde{\bar{J}}_{+}(0)\tilde{\bar{J}}_{+}(\tau)\tilde{\bar{J}}_{-}(\tau)\tilde{\bar{J}}_{-}(0)}_{\rm ss}\\
&=\left(\braket{\tilde{\bar{J}}_{+}}_{\rm ss}\braket{\tilde{\bar{J}}_{-}}_{\rm ss} + \braket{\Delta\tilde{\bar{J}}_{+}\Delta\tilde{\bar{J}}_{-}}_{\rm ss}\right)^2 \\
&+ \braket{\tilde{\bar{J}}_{+}}_{\rm ss}\braket{\tilde{\bar{J}}_{-}}_{\rm ss} \left[\braket{\Delta\tilde{\bar{J}}_{+}(0)\Delta\tilde{\bar{J}}_{-}(\tau)}_{\rm ss} + \text{c.c.}\right]\\
&+ \left(\braket{\tilde{\bar{J}}_{-}}_{\rm ss}\right)^2 \braket{\Delta\tilde{\bar{J}}_{+}(0)\Delta\tilde{\bar{J}}_{+}(\tau)}_{\rm ss} + \text{c.c.}\\
&= \left[X^2 + N^{-1}C_{\rm ss}^{\tilde{\nu}_{*}\tilde{\nu}}(0)\right]^2 + 2N^{-1}X^2 [C_{\rm ss}^{\tilde{\nu}_{*}\tilde{\nu}}(\tau)+C_{\rm ss}^{\tilde{\nu}_{*}\tilde{\nu}_{*}}(\tau)],
\end{aligned}
\end{equation}
where in this approximation we have neglected the fluctuation term $\braket{\Delta\tilde{\bar{J}}_{+}(0)\Delta\tilde{\bar{J}}_{+}(\tau)\Delta\tilde{\bar{J}}_{-}(\tau)\Delta\tilde{\bar{J}}_{-}(0)}_{\rm ss}$ which is of second order in $N^{-1}$, and we have used the fact that third-order terms in the fluctuations vanish when averaged by a Gaussian distribution in the linearized analysis about the steady state of optical bistability. 
\begin{figure}
\centering
\includegraphics[width=0.45\textwidth]{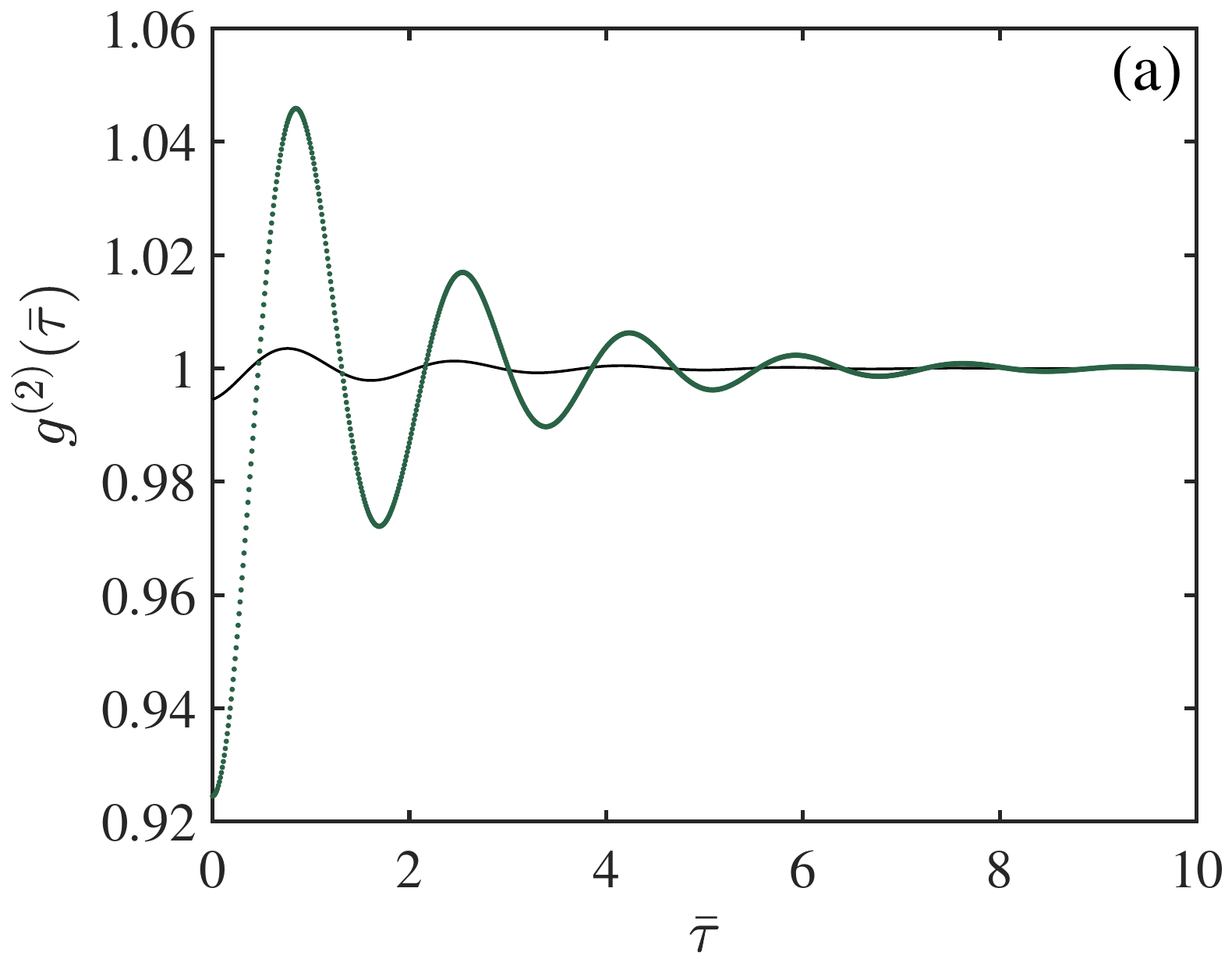}
\includegraphics[width=0.45\textwidth]{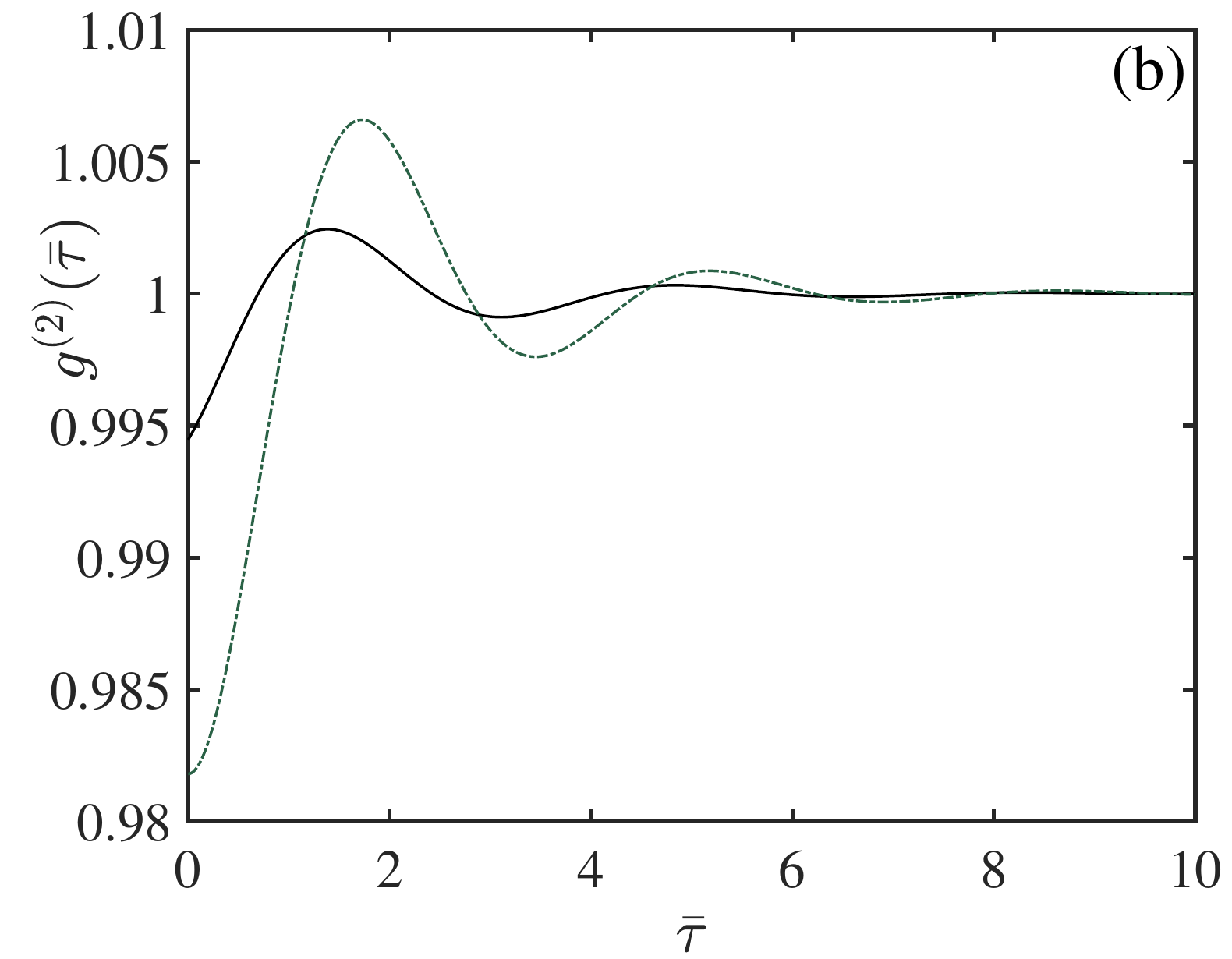}
\caption{{\it Intensity correlations for the collective atomic polarization in the weak-excitation limit, against the reduced time delay $\bar{\tau}=\gamma\tau/2$.} {\bf (a)} The second-order correlation function $g^{(2)}(\bar{\tau})$ from Eq. \eqref{eq:g2WE_B} is plotted in solid black on top of the second-order correlation function of the forwards-scattered field, $g^{(2)}_{\rm F}(\bar{\tau})$, from Eq. \eqref{eq:g2FSc} in green dots, for $(g,\kappa, \gamma)/2\pi=(1.06, 0.88, 10)\,$MHz, $N=310$ [parameters used in the experiment by Raizen and coworkers \cite{Raizen1989} to give $C \approx 40$ for the typical hierarchy of scales $g\sqrt{N} > \gamma > 2\kappa$.]. {\bf (b)} The correlation function $g^{(2)}(\bar{\tau})$ from Eq. \eqref{eq:g2WE_B} is plotted in solid black on top of  $g^{(2)}_{\rm F}(\bar{\tau})$ from Eq. \eqref{eq:g2FSc} in a dotdash green line, for the same parameters as those used in frame (a) except for the coupling strength which is instead lowered to a half of the value used in (a), namely $g/2\pi=0.53\,$MHz, leading to $C \approx 10$.}
\label{fig:intcorrelations}
\end{figure}

\subsection{Intensity correlations along the cooperative branch}
\label{subsec:WELG2}

In the region of very weak intracavity excitation along the lower branch of the bistability curve, $X \ll X_{-}$, we approximate
\begin{equation}
\braket{\tilde{\bar{J}}_{+}\tilde{\bar{J}}_{-}}_{\rm ss}\approx X^2 + N^{-1}C_{\rm ss}^{\tilde{\nu}_{*}\tilde{\nu}}(0).
\end{equation}
Keeping dominant terms in $X^2$ and $N^{-1}$, which amounts to neglecting the contribution from $C_{\rm ss}^{\tilde{\nu}_{*}\tilde{\nu}}(\tau) \sim X^4$, we obtain
\begin{equation}\label{eq:g2WE_A}
g^{(2)}(\tau)=1+2N^{-1}\frac{C_{\rm ss}^{\tilde{\nu}_{*}\tilde{\nu}_{*}}(\tau)}{X^2}.
\end{equation}
Neglecting the standard correlation $C_{\rm ss}^{\tilde{\nu}_{*}\tilde{\nu}}(\tau)$ in favor of the anomalous correlation $C_{\rm ss}^{\tilde{\nu}_{*}\tilde{\nu}_{*}}(\tau) \sim X^2$ in the weak-excitation limit is indicative of the nonclassicality in the fluctuations along the lower branch, and is also observed in the second-order correlation function of the forward-scattered light. From Eq. \eqref{eq:firsttwo_b}, we write
\begin{equation}
\begin{aligned}
&-\frac{(\xi+1)(1+2C)}{X^2}\bar{\mathcal{C}}_{\rm ss}^{\tilde{\nu}_{*}\tilde{\nu}_{*}}(\bar{s})=\frac{(1+\xi+2C)\bar{s}+\xi(\xi+1)}{(\xi +\bar{s})(1+\bar{s})+\xi 2C}\\
&=\frac{1}{\bar{\lambda}_{+}-\bar{\lambda}_{-}}\Bigg[\xi(\xi+1)\left(\frac{1}{\bar{s}-\bar{\lambda}_{+}}-\frac{1}{\bar{s}-\bar{\lambda}_{-}} \right)\\
& + (1+\xi+2C)\left(\frac{\bar{\lambda}_{+}}{\bar{s}-\bar{\lambda}_{+}}-\frac{\bar{\lambda}_{-}}{\bar{s}-\bar{\lambda}_{-}} \right) \Bigg],
\end{aligned}
\end{equation}
whence, after calculating the inverse Laplace transform we obtain the correlator
\begin{equation}\label{eq:CorrInv}
\begin{aligned}
&C_{\rm ss}^{\tilde{\nu}_{*}\tilde{\nu}_{*}}(\bar{\tau})=-\frac{X^2}{(\xi+1)(1+2C)} e^{-\frac{1}{2}(\xi+1)\bar{\tau}}\\
&\left[(1+\xi+2C)\cos(\bar{G}\bar{\tau})+\frac{(\xi+1)(\xi-1-2C)}{2\bar{G}}\sin(\bar{G}\bar{\tau}) \right].
\end{aligned}
\end{equation}
In Eq. \eqref{eq:CorrInv}, $\bar{G}\equiv \sqrt{\xi 2C-\frac{1}{4}(\xi-1)^2}$ is the scaled frequency of the vacuum Rabi oscillation \cite{Carmichael1986} which we met earlier in Eq. \eqref{eq:eigenlambda}, an expression which also applies to the single-atom case $N=1$ [we note that $G\equiv(\gamma/2)\bar{G}=\sqrt{N}g$ for impedance matching, $\gamma=2\kappa$]. Finally, substituting in Eq. \eqref{eq:g2WE_A}, we get the following expression for the {\it second-order correlation function of collective atomic polarization in the weak-excitation limit of absorptive bistability}:
\begin{equation}\label{eq:g2WE_B}
\begin{aligned}
&g^{(2)}(\bar{\tau})=1-\frac{2(1+\xi+2C)}{N(\xi+1)(1+2C)} e^{-\frac{1}{2}(\xi+1)\bar{\tau}}\\
&\times\left[\cos(\bar{G}\bar{\tau})+\frac{(\xi+1)(\xi-1-2C)}{2\bar{G}(1+\xi+2C)}\sin(\bar{G}\bar{\tau}) \right].
\end{aligned}
\end{equation}
In the good-cavity limit ($\xi \ll 1$), the size of the nonclassical fluctuation $g^{(2)}(0)-1 \approx -2 N^{-1}$ is indicative of the number of atoms in the cavity; in the bad-cavity limit ($\xi \gg 2C$), this deviation is further compromised by the large atomic cooperativity $2C$. The expression of Eq. \eqref{eq:g2WE_B} can be recast in the following form:
\begin{equation}\label{eq:g2alternWE}
\begin{aligned}
&g^{(2)}(\tau)=1-\frac{2(\kappa+\gamma^{\prime}/2)}{N(\kappa+\gamma/2)(1+2C)} e^{-\frac{1}{2}(\kappa+\gamma/2)\tau}\\
&\times \left[\cos(g^{\prime} \tau)+\frac{r(\kappa, \gamma, 2C)}{g^{\prime}}\sin(g^{\prime} \tau)\right],
\end{aligned}
\end{equation}
where we have defined the effective impedance
\begin{equation*}
r(\kappa, \gamma, 2C) \equiv \frac{1}{2}(\kappa+\gamma/2)\frac{\kappa-\gamma^{\prime}/2}{\kappa+\gamma^{\prime}/2},
\end{equation*}
with $g^{\prime} \equiv \sqrt{Ng^2-[(1/2)(\kappa-\gamma/2)]^2}$, $\gamma^{\prime} \equiv \gamma(1+2C)$, and $2C=2Ng^2/(\kappa\gamma)$ featuring explicitly as the collective spontaneous emission enhancement factor. We compare with Eq. (16.65) of \cite{QO2} giving the second-order correlation function of the side-scattered light obtained from a pure-state factorization \cite{Carmichael1991OC} for a single atom ($N=1$),
\begin{equation}\label{eq:g2sqfull}
\begin{aligned}
&g^{(2)}(\tau)=\\
&\Bigg\{1 - e^{-\frac{1}{2}(\kappa+\gamma/2)\tau} \left[\cos(g^{\prime} \tau)+\frac{r(\kappa, \gamma, 2C)}{g^{\prime}}\sin(g^{\prime} \tau)\right] \Bigg\}^2.
\end{aligned}
\end{equation}
It becomes explicit that neglecting second-order terms in $N^{-2}$ in the derivation of Eq. \eqref{eq:g2WE_B} amounts to missing the third term in the expansion of the square in Eq. \eqref{eq:g2sqfull} that always produces $g^{(2)}(0)=0$ for $N=1$  (as expected from a single atom) setting aside the difference in the coefficient of the middle term in the expansion of the square \textemdash{the} difference in the two prefactors can be neglected in the bad-cavity limit and in the absence of bistability, formally for $\xi \to \infty, C \to 0$. Moreover, overlooking for the moment the constraints imposed by the small-noise analysis, one requires $[2g^2/(\kappa\gamma)]>[(\kappa+\gamma/2)/(\kappa-\gamma/2)]$ for Eq. \eqref{eq:g2alternWE} to be admissible and produce a non-negative intensity correlation; this condition is certainly not satisfied for $\kappa < \gamma/2$, as was used in Fig. \ref{fig:intcorrelations}, for instance, and in the experiment of \cite{Raizen1989}, precluding $N=1$ altogether. This example explicitly demonstrates the limitations of the linear theory of quantum fluctuations. It is also instructive to contrast Eqs. \eqref{eq:g2WE_B}, \eqref{eq:g2alternWE}, with the expression of the correlation function for the side scattered light when the cooperativity parameter $2C$ is large but $2C/N$ (i.e., the single-atom cooperativity) is negligible [see Eq. (16.69) of \cite{QO2} and Sec. 2.3.4 of \cite{QO1}],
\begin{equation}\label{eq:g2side2C}
g^{(2)}(\tau)=1+X^2[2e^{-\overline{\tau}}-e^{-2\overline{\tau}}],
\end{equation}
which predicts weak photon bunching, $g^{(2)}(0)=1+X^2$, and never drops below unity. For the derivation of Eq. \eqref{eq:g2side2C} we use the fact that the ratio of unlike-atom to like-atom correlations is of order $N^{-1}$ as well as \textemdash{in} a frame rotating with $\omega_0$ \textemdash{the} approximation $\braket{\Delta\tilde{\sigma}_{+}(0)\Delta\tilde{\sigma}_{-}(\tau)}_{\rm ss} \approx X^4 [e^{-\overline{\tau}}-(1/2)e^{-2\overline{\tau}}]$ in the weak-drive limit of resonance fluorescence (the intracavity-field amplitude replaces here the drive strength in free-space resonance fluorescence).

Let us now consider the second-order correlation function of forwards photon scattering,
\begin{equation}\label{eq:g2FSc}
\begin{aligned}
&g^{(2)}_{\rm F}(\bar{\tau})=1-\frac{2}{N}\left(\frac{\xi}{\xi+1}\right) \left(\frac{4C^2}{1+2C}\right) e^{-\frac{1}{2}(\xi+1)\bar{\tau}}\\
&\times \left[\cos(\bar{G}\bar{\tau})+\frac{\xi+1}{2\bar{G}}\sin(\bar{G}\bar{\tau}) \right],
\end{aligned}
\end{equation}
which predicts antibunching along the initial part of the lower branch of absorptive bistability in the bad-cavity limit \cite{Casagrande1980, Carmichael1983}. The intensity correlation function of atomic polarization $g^{(2)}(\bar{\tau})$, from Eq. \eqref{eq:g2WE_B}, is compared to $g^{(2)}_{\rm F}(\bar{\tau})$ from Eq. \eqref{eq:g2FSc} in Fig. \ref{fig:intcorrelations}(a) for the parameters used in the experiment of \cite{Raizen1989}. Lowering the atom-field coupling strength to half of its initial value, and consequently decreasing substantially the cooperativity, reduces the relative deviation from unity for the two correlators, as we observe in Fig. \ref{fig:intcorrelations}(b): the magnitude of the fluctuation $|g^{(2)}(\tau)-1|$ remains fixed to the value $2 N^{-1}$ while photon antibunching for the forwards scattered light is weaker due to a decrease in $2C$. In the weak-excitation limit, the negative source-field spectrum of squeezing can be extracted via the auxiliary output channel, introduced in Sec. \ref{sec:accessatem}, as proportional to the dominant contribution of ${\rm Re}[\bar{\mathcal{C}}_{\rm ss}^{\tilde{\nu}_{*}\tilde{\nu}_{*}}(\bar{s})]$ read from Eq. \eqref{eq:firsttwo_b} with $\bar{s}=-2i\omega/\gamma$. This is accomplished via homodyne detection for a local oscillator in phase with the mean atomic polarization. The negative spectrum of squeezing and antibunching for the auxiliary output field of collective emission are then both linked to the same anomalous correlation of order $X^2$, as it happens for the forwards-scattered light.  
 
Specializing now for simplicity to the case of impedance matching, $\kappa=\gamma/2$ ($\xi=1$) \textemdash{frequently} encountered in the literature \textemdash{we} obtain
\begin{equation}\label{eq:g2IM}
\begin{aligned}
&g^{(2)}(\tau)-1=-\frac{1}{N(1+2C)} e^{-\kappa\tau}\\
&\times \left[2(1+C)\cos(\sqrt{N}g\tau)-\sqrt{2C}\,\sin(\sqrt{N}g\tau) \right].
\end{aligned}
\end{equation}
As for the forwards-scattered light for $\kappa=\gamma/2$,
\begin{equation}\label{eq:g2FIM}
\begin{aligned}
&g^{(2)}_{\rm F}(\tau)-1=-\frac{4C^2}{N(1+2C)} e^{-\kappa\tau}\\
&\times\left[\cos(\sqrt{N}g\tau)+(1/\sqrt{2C})\sin(\sqrt{N}g\tau)\right].
\end{aligned}
\end{equation}
We note that for a dominant cavity-emission enhancement, the right-hand sides of Eqs \eqref{eq:g2IM} and \eqref{eq:g2FIM} differ primarily by a factor of $2C$ which sets the ratio between the total spontaneous emission rate and the rate of cavity emissions through the mirrors of the resonator (see Note 16.4 of \cite{QO2}). As the collective light-matter coupling strength decreases, the two intensity correlations oscillate out of phase with each other. 

\subsection{One-atom behavior along the independent branch}
\label{subsec:SEG2}

In the regime where $X \gg X_{+}$, the steady-state collective polarization is $\braket{\tilde{\bar{J}}_{+}}_{\rm ss}=\braket{\tilde{\bar{J}}_{-}}_{\rm ss} \approx 1/X$, while $\braket{\Delta\tilde{\bar{J}}_{+}\Delta\tilde{\bar{J}}_{-}}_{\rm ss} \approx 1/N$. Both these quantities are considered through their ratio albeit very small in their own right. This means that the intracavity excitation is to be compared against the system-size parameter. After inverting the Laplace transform of
\begin{equation}
\begin{aligned}
&\bar{\mathcal{C}}_{\rm ss}^{\tilde{\nu}_{*}\tilde{\nu}}(\bar{s})+\bar{\mathcal{C}}_{\rm ss}^{\tilde{\nu}_{*}\tilde{\nu}_{*}}(\bar{s})=\frac{\bar{s}+2}{2X^2+(\bar{s}+1)(\bar{s}+2)}=\frac{1}{\bar{\rho}_{+}-\bar{\rho}_{-}}\\
&\times \left[2\left(\frac{1}{\bar{s}-\bar{\rho}_{+}}- \frac{1}{\bar{s}-\bar{\rho}_{-}}\right)+\left(\frac{\bar{\rho}_{+}}{\bar{s}-\bar{\rho}_{+}}- \frac{\bar{\rho}_{-}}{\bar{s}-\bar{\rho}_{-}}\right) \right],
\end{aligned}
\end{equation}
where $\bar{\rho}_{\pm}=-3/2 \pm i \sqrt{2}X$, and using Eqs. \eqref{eq:defg2} and \eqref{eq:numg2} with the appropriate steady-state averages, we obtain the strong-excitation counterpart of Eq. \eqref{eq:g2WE_B}:
\begin{equation}\label{eq:g2SC}
\begin{aligned}
&g^{(2)}(\bar{\tau})-1=\frac{2NX^2}{(N+X^2)^2}[C_{\rm ss}^{\tilde{\nu}_{*}\tilde{\nu}}(\bar{\tau})+C_{\rm ss}^{\tilde{\nu}_{*}\tilde{\nu}_{*}}(\bar{\tau})]\\
&=\frac{2NX^2}{(N+X^2)^2} e^{-3\bar{\tau}/2}\left[\cos(\sqrt{2}X \bar{\tau})+\frac{1}{2\sqrt{2}}\sin(\sqrt{2}X \bar{\tau})\right].
\end{aligned}
\end{equation}
We note that $g^{(2)}(0)>1$ for every value of the intracavity excitation in this limit; the linearized treatment of fluctuations imposes the constraint $X^2 \ll N$ yielding a prefactor $[2NX^2/(N+X^2)^2]\approx 2X^2/N \ll 1$. In the expression for the intensity correlation function, terms of order $N^{-2}$ may still be neglected. In Eq. \eqref{eq:g2SC}, the two correlation functions $C_{\rm ss}^{\tilde{\nu}_{*}\tilde{\nu}}(\bar{\tau})$ and $C_{\rm ss}^{\tilde{\nu}_{*}\tilde{\nu}_{*}}(\bar{\tau})$ contribute by the same order of magnitude to the final result, unlike in the correlation function of Eq. \eqref{eq:g2WE_B} in the weak-excitation limit. Once more, similarly to the spectrum of Eq. \eqref{eq:TSELcoll}, this expression does not depend on $C$ or $\xi$, while the number of atoms $N$ only sets the upper boundary for the intracavity excitation to determine the size of the fluctuation $[g^{(2)}(0)-1]$.

\section{Concluding discussion}
\label{sec:conclusions}

In this study, we have investigated the small-noise incoherent correlation spectrum of absorptive optical bistability alongside squeezing of fluctuations and the intensity correlation function of the sideways-scattered field without recourse to the adiabatic elimination of any system variables. This was done in the weak-excitation limit, where analytical expressions can be obtained in a consistent way via the expansion to lowest order of the equations of motion for the covariance matrix and the initial conditions, and in the high excitation region with apparent similarities to free-space resonance fluorescence. Furthermore, we have proposed a setup employing an auxiliary low-Q cavity to translate the atomic correlations to a measurable output, since collective effects are suppressed in the side scattering from distinguishable atoms with individual detectable records. Adiabatically eliminating the auxiliary cavity field sets up a collective emission channel through which the atomic fluorescent spectrum can be imaged with the inclusion of unlike-atom correlations.

We have seen that, in the weak-excitation limit, the collective atomic polarization correlations \textemdash{corresponding} to the fluctuations of an internal system degree of freedom \textemdash{follow} those of the forwards-scattered field which is readily accessible by the experiment in a single collective mode. We have focused on three limiting regimes defined by the ratio of the effective coupling strength to the dissipation rates as well as by the ratio between those rates themselves. The presence of squeezing is the element reinforcing the correspondence between the two coupled degrees of freedom, whose fluctuations have spectral densities exhibiting a $|\omega-\omega_0|^{-4}$ dependence for large frequencies. Along the upper branch of the bistability curve with quasi unitary slope, however, the forwards-scattered field and the atomic polarization have differing spectral distributions. The latter exhibits a dynamical Stark shift depending only on the scaled intracavity amplitude, a fact marking the absence of cooperation between the two-level emitters. The former is a Lorentzian having a width which is solely determined by dissipation. 

Lastly, let us briefly comment on the relevance of the regime of optical bistability under consideration to current investigations on cavity-mediated atomic coherence. External manipulation of bistability and the associated critical slowing down underlies numerous proposals in optical switching since the late 1970s \cite{HGibbs}. Weak pulses of light can be used to realize an all-optical switch where the detuning of a few-photon probe controls a more intense output beam. Based on the phenomenon of recoil-induced resonance, the backaction of optical bistability was captured from the coherent interaction between weak light fields and the collective motion of a strongly dispersive atomic gas of about $5\times 10^8$ Rb atoms in the experiment of \cite{Vengalattore2008}. A year earlier, the access of both branches of dispersive optical bistability, occurring at photon numbers below unity due to the feedback from optical forces exerted in an ensemble of $10^5$ Rb atoms with long-lived coherence, had been reported by \cite{Gupta2007}. In the proposed optical-switch configuration of \cite{Yang2011}, a BEC of more than $10^4$ atoms is confined in a high-finesse optical cavity driven by two fields, following the experiment of \cite{Ritter2009} where a single coherent field drives the cavity mode (see also Fig. 3 of \cite{Zhang2009}). The transverse driven mode employed in \cite{Yang2011} is used to control optical bistability while the longitudinal intracavity field, driven with a significant detuning from the atomic resonance, follows adiabatically the condensate since the photon-loss rate dominates. An analogy can be drawn with to the setup proposed in Sec. \ref{sec:accessatem} for accessing the collective atomic emission, where the auxiliary mode with very weak photon flux follows the atomic ensemble. For the optical switch transversely driven on resonance with the cavity frequency, the steady-state amplitude of the longitudinal cavity field is given by the expression (see Eq. 4 of \cite{Yang2011})
\begin{equation}\label{eq:alphaBEC}
\alpha=\frac{\varepsilon_{d,\parallel}-i\varepsilon_{d,\perp} \displaystyle\int dx |\psi(x; \alpha + \alpha^{*}, |\alpha|^2)|^2 \cos(kx)}{\kappa + i U_0  \displaystyle\int dx |\psi(x;\alpha + \alpha^{*}, |\alpha|^2)|^2 \cos(kx)},
\end{equation}
where $\psi(x; \alpha + \alpha^{*}, |\alpha|^2)$ is the condensate wavefunction (the dependence on $\alpha+\alpha^{*}$ originates from the cavity drive), $U_0$ is the depth of the $\cos^2(kx)$ standing wave potential generated by the atom-light interaction \textemdash{a} function of the coupling strength $g$ \textemdash{times} the mean-field intracavity excitation $|\alpha|^2$, and $\varepsilon_{d,(\parallel,\perp)}$ are the two drive-field amplitudes.  

Measurements along the upper branch of the bistability curve with a high intracavity photon number have been recently reported in \cite{Gothe2019} for a collection of about $1.5 \times 10^5$ thermal ytterbium atoms, marking a resurgence of interest in the nonlinearity arising from the coupling of cold trapped atoms to an optical cavity since the exemplary demonstration of strong bistability for laser cooled and trapped cesium atoms in 1995 \cite{Lambrecht1995}. The dynamics for a system with a steady state described by Eq. \eqref{eq:alphaBEC} is considerably more involved than what is described by ME \eqref{eq:MEbist}, let alone the inherent spatial dependence of the coupling strength. Atoms in a BEC occupy a single mode of the matter-wave field with displaying macroscopic coherence; in that respect they cannot be perceived as distinguishable scatterers. The decoherence leading to individual scattering records on the one hand and the cooperation between the individual emitters mediated by a common channel provided by the intracavity field on the other hand sets up a meaningful basis for the comparison of an atomic ensemble subject to ME \eqref{eq:MEbist} and an atomic ensemble forming a BEC coupled to a cavity mode. In the former, unlike-atom correlations do not add up constructively to a measurable output whereas in the latter, the ensemble is described by a wavefunction and couples to the cavity mode as a single `superatom' with an enhanced strength. We note here that in the weak-excitation limit, the cooperativity of the BEC is a function of $\tilde{g}\sqrt{N}$, where $\tilde{g}$ is the average of the position-dependent single-atom coupling strength against the atomic density.

In the experiment of \cite{Brennecke2007}, the atomic ensemble oscillates between its ground state and a symmetric excited state where a single excitation is shared by all the atoms. This instance calls for the employment of symmetrized states in the pure-state factorization for many atoms in the cavity as has been done within the frame of a perturbative expansion of the density-matrix equations of motion in powers of the drive amplitude \textemdash{excursions} out of the manifold of symmetrized states due to spontaneous emission are neglected (see Sec. 16.1.2 of \cite{QO2} and references therein). Here we meet with the difficulty of employing an enormous Hilbert space of dimension $2^N$, which is required when dealing with spatial effects (see Sec. 16.2 and Introduction to Sec. 16.3 of \cite{QO2}). The repulsive interactions between the atoms in the BEC will influence the onset of bistability which will in turn impact on the spectral distribution. Upon accessing the upper branch in the good-cavity limit, for example, do we then expect the incoherent spectrum of the transmitted light to transition from a spectral hole to a sharp Lorentzian distribution? Will the spectrum of collective atomic emission extracted via the auxiliary mode show the familiar Stark triplet? Questions of the kind point to the role of the collective modes as probes of the criticality exhibited by macroscopic dissipative systems; in this work we have focused on the extraction of collective atomic coherence against a background of unlike-atom correlations which do not add up constructively when collecting the side-scattering record. 

\begin{acknowledgments}
I am grateful to Prof. H. J. Carmichael for instructive discussions and guidance. I also wish to thank C. Lled\'{o} for his assistance in producing Fig. 2. Financial support by the Swedish Research Council (VR) alongside the Knut and Alice Wallenberg foundation (KAW) is as well acknowledged.
\end{acknowledgments}

\bibliography{bibliography}

\begin{center}
 *****
\end{center}

\end{document}